%% file: main.tex
\documentclass[review]{elsarticle}

\usepackage{lineno,hyperref}
\graphicspath{ {Figuras/} }
\usepackage[export]{adjustbox}
\usepackage{float}
\usepackage{amsmath}
\usepackage{xcolor}
\usepackage{multirow}

\usepackage{booktabs}

\usepackage{tabularx}
\usepackage{verbatim} 
\usepackage[per-mode=symbol]{siunitx}

\usepackage{vmargin}
\setpapersize{A4}
\setmargins{2.5cm}     
{1cm}                  
{16cm}                 
{25cm}                 
{10pt}                 
{1cm}                  
{0pt}                  
{1.5cm}                

\makeatletter
\def\@author#1{\g@addto@macro\elsauthors{\normalsize%
    \def\baselinestretch{1}%
    \upshape\authorsep#1\unskip\textsuperscript{%
      \ifx\@fnmark\@empty\else\unskip\sep\@fnmark\let\sep=,\fi
      \ifx\@corref\@empty\else\unskip\sep\@corref\let\sep=,\fi
      }%
    \def\authorsep{\unskip,\space}%
    \global\let\@fnmark\@empty
    \global\let\@corref\@empty  
    \global\let\sep\@empty}%
    \@eadauthor={#1}
}
\makeatother

\usepackage[justification=centering]{caption}
\usepackage{subcaption}												
\begin{document}


\begin{frontmatter}

\title{Estimation of the trajectory and attitude of railway vehicles using inertial sensors with application to track geometry measurement}


\author{J. Gonz\'alez-Carbajal  \fnref{Pedro_address}}
\author{Pedro Urda          \fnref{Pedro_address}}
\author{Sergio Mu\~noz      \fnref{Sergio_address}}
\author{Jos\'e L. Escalona\fnref{Pedro_address}\corref{cor1}}

\cortext[cor1]{Corresponding author: escalona@us.es}

\address[Pedro_address]{Department of Mechanical and Manufacturing Engineering, University of Seville, Spain}
\address[Sergio_address]{Department of Materials and Transportation Engineering, University of Seville, Spain}


\begin{abstract}
This paper describes a novel method for the estimation of the trajectory curve and orientation of a rigid body moving along a railway track. Compared to other recent developments in the literature, the presented approach has the significant advantage of tracking the position and orientation of a railway vehicle using inertial sensors only (a 3D gyroscope and a 3D accelerometer), excluding global position sensors (GNSS or total station) and also excluding global orientation sensors (magnetometers or inclinometers). The algorithm is based on a kinematic model of the relative motion of the body with respect to the track. This kinematic model is used as the system equations of a Kalman filter algorithm that includes in the state vector the coordinates used to define the position and orientation of the body. Two different Kalman filter approaches are described. In the first one, the position and orientation are calculated independently. In the second one, both position and orientation are calculated as a coupled problem. Crucial to the success of the results of the Kalman filters is the use of the correct covariance matrices associated with the system process and the measurements. Again, two approaches are used for the estimation of the covariance matrices. One is based on the use of experimental results with a known output. The other one is based on \textit{constrained maximum likelihood estimation}. 
The calculated trajectory and orientation are applied in this research to the problem of track geometry measurement. This is a very demanding application because the measurement of track irregularities requires millimetric accuracy. A scale track with known design geometry and irregularities is used to conduct experiments for tuning and evaluating the quality of the output of the algorithm. Results show that the developed algorithm is accurate enough for this application. They also show that, using either of the two proposed Kalman filtering approaches, the constrained maximum likelihood method for the estimation of the covariance matrices performs similarly to the known-output method. This is very convenient because it allows a straightforward application of the observation model in different scenarios.
\end{abstract}



\end{frontmatter}


\input{A_Introduction}

\input{B_KinematicModels}

\input{C_KalmanFilters}

\input{D_EstimationCovariances}

\input{E_Experiments}

\input{F_Results} 

\input{G_Conclusions}

\section*{Acknowledgements}
This research was supported by the Spanish \textit{Ministerio de Ciencia e Innovación}, under the program "Proyectos I+D+I–2020", with project reference PID2020-117614RB-I00. This support is gratefully acknowledged.

\bibliography{references}

\input{H_Appendix}

\end{document}

%% file: A_Introduction.tex
\section{Introduction}
\label{sec:introduction}

The problem of finding the trajectory of a body in space using inertial sensors is a classic problem that has been treated by many scientists in the fields of inertial navigation of vehicles, robotics, or biomechanics. In fact, the first application of the Kalman filter in the 60's of the last century was the inertial navigation of spacecrafts in the Apolo project. The inherent difficulty of the estimation of a trajectory is that inertial sensors provide signals that are functions of the time-derivatives (first and second derivatives) of the variables needed to calculate it. Due to the existence of noise in the sensors' signals, the calculated trajectory necessarily drifts unbounded with respect to the real trajectory. This problem can be fixed including new sensors that provide signals proportional of the trajectory coordinates themselves, not their derivatives. To this end, the most popular technology is the \textit{Global Navigation Satellite System}, GNSS. \\

Nowadays there exist numerous algorithms for inertial navigation that work reasonably well in many applications. The use of \textit{state observers} aims to determine the state of the system, which may include the sensors biases \cite{Ahmad2019,Bryne2016,Hua2010} or IMU errors parameters  \cite{Scholte2019}. The \textit{fusion algorithms} are based on combining the information of two or more sensors measuring different variables of the same system, leading to estimated trajectory and orientation without the drift problem. These fusion algorithms can be based on Linear Kalman Filter \cite{Aldimirov2018}, Non-linear Extended Kalman Filter \cite{Vaganay1994,Anderle2018}, Unscented \cite{Kada2016}, Cubature Kalman Filter \cite{Benzerrouk2016}), Complementary Filter \cite{Huang2018} or Unknown Input Filter \cite{Abolhasani2018}. The \textit{optimization algorithms} are based on minimizing the error function between the real orientation (unknown) and the estimations based on measured data. This minimization can be approached by different algorithms such as the Levenberg-Marquardt Algorithm (LMA) \cite{Li2019}, Gauss-Newton Algorithm (GNA), Gradient Descent Algorithm (GDA) \cite{madgwick2010,madgwick2011} and Control Loop \cite{Mansoor2019,Alqaisi2020}. All these algorithms are successfully applied to the trajectory estimation of aircrafts, spacecrafts, cars, robots or humans. The estimation of the trajectory of an aircraft in space is much more difficult than the estimation of the trajectory of a vehicle along a rail track. The obvious reason is that the aircraft moves freely in an unbounded space while the rail vehicle follows more or less closely the track centerline 3D curve. In other words, the trajectory of the rail vehicle is approximately known beforehand. Only slight deviations about the reference trajectory are expected. However, the estimation of the trajectory of a rail vehicle has other difficulties. In order to get advantage of the knowledge of the approximate solution (track centerline), the use of track-relative kinematics is recommended. Track-relative kinematic equations are much more involved than the kinematics of a free body in space. Besides, it requires the use of the design geometry of the track and an odometry algorithm that estimates the position of the vehicle along the track. When the estimation of trajectory of the rail vehicle is used for the estimation of the real track geometry (rail geometry measurement), the accuracy that is needed is millimetric. This is much more than the accuracy needed for the inertial navigation of other vehicles. This paper treats the problem of the trajectory estimation of railways with application to track geometry measurement.\\  

In the railway industry, track geometry measurement is a fundamental phase of track maintenance. Two different types of equipment are used for this task: man-driven \textit{rail track trolleys} (RTT) and \textit{track recording vehicles} (TRV). On one hand, the technology behind the RTT is simple and accurate. The relative irregularities (track gauge and cross-level) are directly measured with a distance sensor, like an LVDT and an inclinometer, respectively. The absolute irregularities (alignment and vertical profile) require and absolute positioning system, like a total station or a very accurate GNSS. Despite of their precision and good performance, in addition to its low cost, the main handicap of RTT devices is their slowness when measuring the track.  On the other hand, the technology used in TRV is more varied. Essentially, there are two technologies \cite{Grassie1996}: versine measurement systems (VMS), and inertial measurement systems (IMS). The VMS, also called chord method, are based on simple kinematics that only requires the measurements of distance sensors, obtaining both horizontal and vertical measurement of rails \cite{Chiou2019}. The IMS are based on the use of inertial sensors (accelerometers and gyroscopes), the most serious problem being the need to integrate the sensor signals in time to get the irregularities. The main drawback of the TRV methods is that are based on the use of expensive and dedicated vehicles.    \\

The alternative to both measuring methods (RTT and TRV) is the use of inexpensive measuring systems, mounted on in-service vehicles for continuous monitoring of track conditions. That way, the frequency of the track inspection is increased and the time-evolution of the irregularities can be followed closely, all this at a very low cost. An extensive review on the use of in-service vehicles for the monitoring of railway tracks can be found in \cite{Weston2015}. Weston et al. published a set of papers\cite{Weston2007, Weston2007b}, showing the measurement of vertical and lateral track irregularities using accelerometers and gyroscopes mounted in the bogie frame and a very simple kinematic model of the vehicle motion.  Lee et al. \cite{ Lee2012} presented a Kalman filter data fusion approach based on accelerometer signals mounted on the body frame and the axle-box. In this work, the Kalman filter is used as a kind of integrator of lateral acceleration of the wheelset to obtain lateral displacements and, subsequently, a set of compensation filters are used to predict lateral irregularities. Wei et al. \cite{Wei2016} proposed a method to estimate the alignment of the track through a double integration of the acceleration measured by several accelerometers placed on the vehicle.  Tsai et al. \cite{Tsai2012, Tsai2014} presented a fast inspection technique based on the Hilbert–Huang Transform, that is a very useful time–frequency analysis technique, applied to the signal of an accelerometer mounted on the axle-box of an in-service car. Both methods that are based on axle-box-mounted accelerometers are only valid to measure vertical track irregularities. Tsumashima et al. \cite{Tsunashima2014} estimated the vertical track geometry using the accelerations measured in the car-body of the Japanese Shinkansen using a Kalman filter based on a very basic car model.  More recently, De Rosa et. al. \cite{DeRosa2019} proposed three different model-based methods to estimate both lateral track alignment and cross-level irregularities: 1) pseudo-inversion of the vehicles frequency response function (FRF) matrix, 2) unknown input estimation using a deterministic observer and 3) unknown input estimation using a linear Kalman filter as a stochastic observer.  In \cite{Munoz2021} Muñoz et. al. proposed a Kalman filter model-based methodology for the estimation of lateral track irregularities from measurements from different sensors mounted on an in- service vehicle. The proposed methodology was experimentally validated through an experimental campaign carried out in a 1:10 scaled vehicle, obtaining promising results. In the work of Escalona et al. \cite{escalona2021}, a track measurement system that can be installed on in-service vehicles and combines a kinematic model, a computational vision system and inertial measurement is presented.

Finally, it can be concluded that, many of the different methodologies for the track geometry measurement require the estimation of the trajectory for the measurement of the absolute irregularities. Consequently, it would be very convenient to develop an accurate method for the calculation of the trajectory of the vehicles using only inertial measurements. In the present work, a new method for the estimation of the trajectory and orientation of a rigid body is presented, with application to track geometry measurement. \\

This paper is organized as follows. Section 2 explains the kinematic description of the geometry of a track, of a vehicle running on the track, and their relationship with the measurements of inertial sensors. Section 3 explains the different Kalman filters used to estimate the trajectory and attitude of a body moving along the track. Section 4 shows two different methods that can be used to find the covariance matrices needed for the application of the Kalman filters. Section 5 describes the experiments used for the application of the developed estimation techniques to the geometry measurement of a scale track. Section 6 shows and discusses the validity of the estimation techniques based on the comparison with the experimental measurements. Finally, summary and conclusions of this work are given in Section 7.

%% file: B_KinematicModels.tex
\section{Kinematic models} \label{sec:KinematicModels}

This section shows the kinematic models used in the estimation of the trajectory and orientation of a body moving along a rail track. It is a summary of the more detailed kinematic description given in \cite{escalona2021}. This section describes the kinematics of the track design geometry, the kinematics of the body based on track-relative kinematics and the relationships between the signals measured with the inertial sensors and the body and track kinematics.

\begin{figure}[htbp!]
	\centering
	\includegraphics[width=0.8\linewidth]{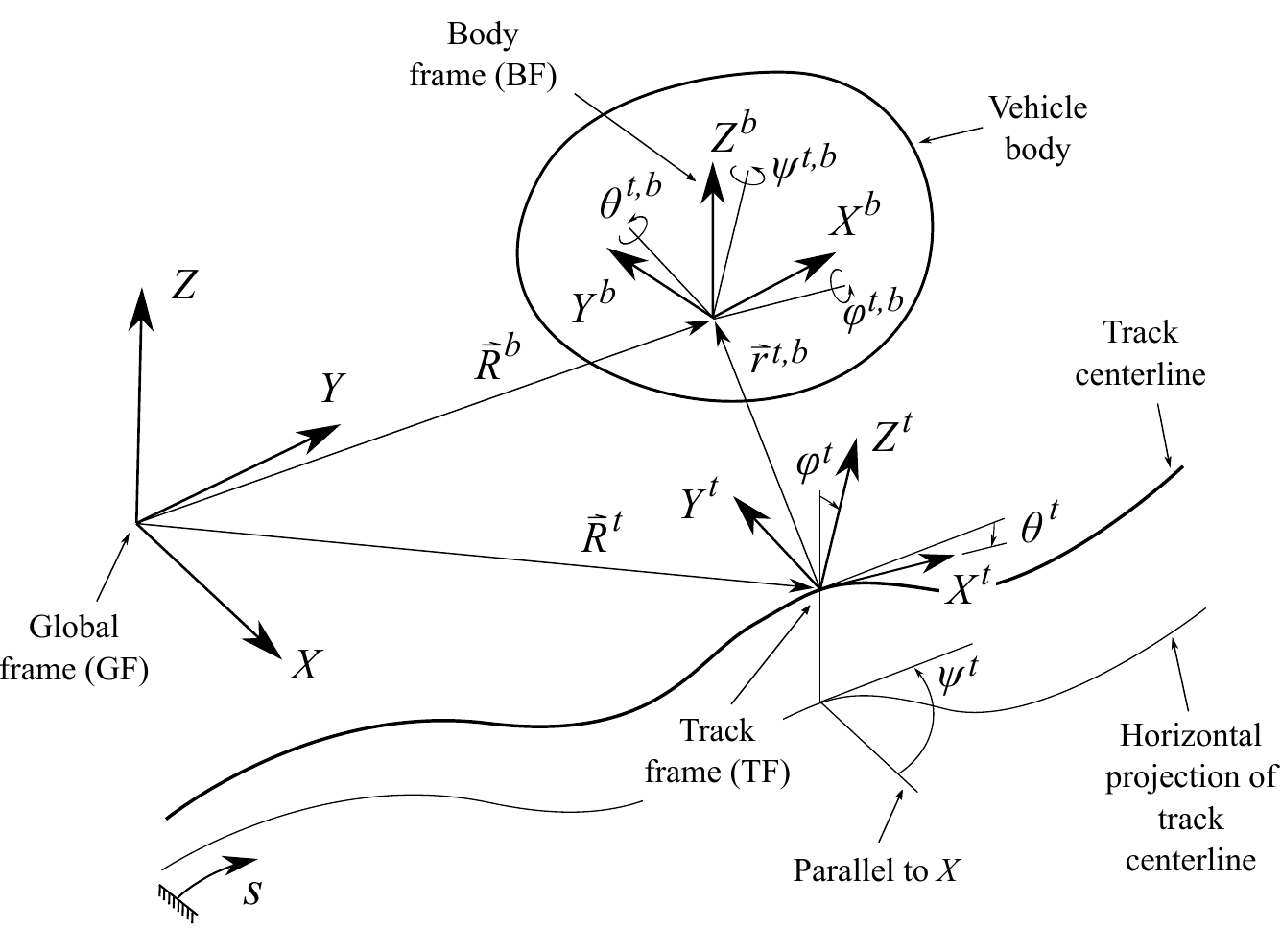}
	\caption{Kinematics of a body with track-relative coordinates}
	\label{fig:railKinematics3D}
\end{figure}

Figure \ref{fig:railKinematics3D} shows an arbitrary rigid body $b$ that moves along a rail track. The track is simply represented by its centerline. The figure shows three frames: (1) an inertial and \textit{global frame} (GF) $<X, Y, Z>$ that is fixed in space, (2) a \textit{track frame} (TF) $<X^t, Y^t, Z^t>$ whose position and orientation are known functions of the arc-length coordinate along the track $s$ and (3) the body frame (BF) $<X^b, Y^b, Z^b>$ that is rigidly attached to the body. 

\subsection{Kinematics of the track centerline} \label{sec:trackCenterline}

Describing the kinematics of the track centerline is equivalent to providing the functions that give the position and orientation of the TF with respect to the GF.
The definition of the TF is such that the $X^t$ axis is tangent to the track design centerline, the $Y^t$ axis is perpendicular to $X^t$ and connects the left rail centerline and the right rail centerline and the $Z^t$ axis is perpendicular to both $X^t$ and $Y^t$. The components of the absolute position vector of an arbitrary point on the design track centerline with respect to the GF is a function of the arc-length $s$, as follows: 

\begin{equation} \label{Eq3_1}
{{\bf{R}}^t}\left( s \right) = \left[ {\begin{array}{*{20}{c}}
{R_x^t}\\
{R_y^t}\\
{R_z^t}
\end{array}} \right]
\end{equation}

And the rotation matrix from the TF to the GF can be approximated to: 

\begin{equation} \label{Eq3_3}
{{\bf{A}}^t}\left( s \right) \simeq \left[ {\begin{array}{*{20}{c}}
{{\rm{c}}{\psi ^t}}&{ - {\rm{s}}{\psi ^t}}&{{\varphi ^t}{\rm{s}}{\psi ^t} + {\theta ^t}{\mathop{\rm c}\nolimits} {\psi ^t}}\\
{{\rm{s}}{\psi ^t}}&{{\mathop{\rm c}\nolimits} {\psi ^t}}&{{\theta ^t}{\rm{s}}{\psi ^t} - {\varphi ^t}{\rm{c}}{\psi ^t}}\\
{ - {\theta ^t}}&{{\varphi ^t}}&1
\end{array}} \right]
\end{equation}

where the Euler angles ${\psi ^{t}}$  ($azimut$ or $heading$ angle), ${\theta ^{\,t}}$  ($vertical slope$) and ${\varphi ^{\,t}}$  ($cant$ or $superelevation$ $angle$) that can be observed in Fig. \ref{fig:railKinematics3D} have been used. In this equation, ``s'' means ``sine'' and ``c'' means ``cosine'' of the angle. Note that the vertical slope and cant angles are assumed to be small and the rotation matrix is linearized using the small-angle assumption. 

The six functions $R_x^t\left( s \right)$, $R_y^t\left( s \right)$, $R_z^t\left( s \right)$, ${\psi ^{t}\left( s \right)}$, ${\theta ^{\,t}\left( s \right)}$, and ${\varphi ^{\,t}\left( s \right)}$ are tabulated in the railway industry and they represent the design geometry of the track centerline. In this paper, it is assumed that these functions are given. 

Each body $b$ moving along the track has an associated TF at each instant of time. Its position and orientation is obtained substituting the position of the body along the track, $s^b(t)$, in the functions ${{\bf{R}}^t}\left( s \right)$ and ${{\bf{A}}^t}\left( s \right)$.\\

\subsection{Kinematics of a body moving along the track} \label{sec:bodyAlongTrack}
The coordinates used to describe the position and orientation of the arbitrary body $b$ shown in Fig. \ref{fig:railKinematics3D} moving along the track are:

\begin{equation} \label{bodyCoord}
{{\bf{q}}^b} = {\left[ {\begin{array}{*{20}{c}}
{{s^b}}&{r_y^{t,b}}&{r_z^{t,b}}&{{\varphi ^{t,b}}}&{{\theta ^{t,b}}}&{{\psi ^{t,b}}}
\end{array}} \right]^T}
\end{equation}

where $s^b$ is the arc-length along the track of the position of the body, $r_y^{t,b}$ and $r_z^{t,b}$ are the non-zero components of the position vector $\vec r^{t,b}$ of the BF with respect to the TF, this is ${{\bf{\bar r}}^{t,b}} = {\left[ {\begin{array}{*{20}{c}}
0&{r_y^{t,b}}&{r_z^{t,b}}
\end{array}} \right]^T}$  , and ${\varphi ^{t,b}},\,\,{\theta ^{t,b}}\,\,{\rm{and}}\,\,{\psi ^{t,b}}$ are three Euler angles (roll, pitch and yaw, respectively) that define the orientation of the BF with respect to the TF. 

The linearized transformation matrix from the BF to the TF is given by:

\begin{equation} \label{Eq3_12}
{{\bf{A}}^{t,b}} \simeq \left[ {\begin{array}{*{20}{c}}
1&{ - {\psi ^{t,b}}}&{{\theta ^{t,b}}}\\
{{\psi ^{t,b}}}&1&{ - {\varphi ^{t,b}}}\\
{ - {\theta ^{t,b}}}&{{\varphi ^{t,b}}}&1
\end{array}} \right]
\end{equation}

The absolute position vector and the absolute orientation matrix of body $b$ are given by:

\begin{equation} \label{Eq3_13}
{\bf R}^b = {\bf R}^t +  {\bf A}^t {{\bf{\bar r}}^{t,b}}, \;  {\bf A}^b = {\bf A}^t  {\bf{A}}^{t,b} 
\end{equation}

The time-derivatives of these expressions are used to find the absolute acceleration and angular velocity of body $b$. See more details in \cite{escalona2021}.

\subsection{Kinematics of the gyroscope signals} \label{sec:gyroKin}

In the remainder of this paper it will be assumed that an inertial sensor or IMU is installed on the moving body. The IMU is assumed to be installed at the origin of the BF of the moving body with the sensor frame parallel to the BF. The IMU measures the 3 components of the absolute angular velocity and the the three components of the absolute acceleration vectors in the sensor frame. The three signals measured with the gyroscope can be interpreted as:

\begin{equation} \label{wImu}
{{\boldsymbol{\omega}}^{imu}} = \hat{ \boldsymbol {\omega}}^{b} 
\end{equation}

where ${{\boldsymbol{\omega}}^{imu}}$ is the $3 \times 1$ array of the gyroscope signals and $\hat{ \boldsymbol {\omega}}^{b} $ contains the 3 components of the absolute angular velocity in the BF. As explained in \cite{escalona2021}, the time derivative of the absolute Euler angles of the body $b$ are related to the gyroscope signals as follows:

\begin{equation} \label{dEulOmega}
\left[ {\begin{array}{*{20}{c}}
{\dot \varphi^{b} }\\
{\dot \theta^{b} }\\
{\dot \psi^{b} }
\end{array}} \right] = \left[ {\begin{array}{*{20}{c}}
1&{\frac{{\sin \varphi^{b} \sin \theta^{b} }}{{\cos \theta^{b} }}}&{\frac{{\cos \varphi^{b} \sin \theta^{b} }}{{\cos \theta^{b} }}}\\
0&{\cos \varphi^{b} }&{ - \sin \varphi^{b} }\\
0&{\frac{{\sin \varphi^{b} }}{{\cos \theta^{b} }}}&{\frac{{\cos \varphi^{b} }}{{\cos \theta^{b} }}}
\end{array}} \right]\left[ {\begin{array}{*{20}{c}}
{{{\hat \omega }_x}}\\
{{{\hat \omega }_y}}\\
{{{\hat \omega }_z}}
\end{array}} \right]
\end{equation}

where ${\bf{\Phi }}^{b} = {[\begin{array}{*{20}{c}}
{\varphi^{b}}&{\theta^{b}}&{\psi^{b}}
\end{array}]^T}$ is the set of absolute angles (with respect to the GF) of the body $b$. Note that these are not the same angles given in Eq. \ref{bodyCoord} which are BF to TF relative angles. This expression is non-linear in terms of absolute Euler angles, but linear in terms of their time-derivatives. Using again the small angles assumption of the roll and pitch angles, the following linearization can be adopted:

\begin{equation} \label{Eq7_3}
\begin{array}{l}
\dot \varphi^{b} \approx {{\hat \omega }_x} + \theta^{b} {{\hat \omega }_z}\\
\dot \theta^{b} \approx {{\hat \omega }_y} - \varphi^{b} {{\hat \omega }_z}\\
\dot \psi^{b}  \approx \varphi^{b} {{\hat \omega }_y} + {{\hat \omega }_z}
\end{array}
\end{equation}

\subsection{Kinematics of the accelerometer signals} \label{sec:accelKin}

Nowadays, most IMU's used in vehicle dynamics have MEMS-type accelerometers. In contrast to piezoelectric accelerometers, the measured signals include the effect of gravity, as follows:

\begin{equation} \label{accelSignals}
{{\bf{a}}^{imu}} = \hat{ \ddot {\bf{R}}}^{b} + {\left( {{{\bf{A}}^{b}}} \right)^T}{\left[ {\begin{array}{*{20}{c}}
0&0&g
\end{array}} \right]^T}
\end{equation}

where ${{\bf{a}}^{imu}}$ is the $3 \times 1$ array of the accelerometer signals, $\hat{ \ddot {\bf R}}^{b}$ contains the 3 components of the absolute acceleration of the IMU in the sensor frame, ${\bf{A}}^{b}$ is the absolute rotation matrix of the IMU and $g$ is the acceleration of gravity that is assumed to act in the $Z$ direction. As shown in \cite{escalona2021}, developing this equation the following approximate relationship between the accelerometer signals and the body coordinates are obtained:
\begin{equation}
\begin{aligned} \label{Eq9_3}
\begin{bmatrix}
0 \\ \ddot{r}_y^{t,b} \\ \ddot{r}_z^{t,b}
\end{bmatrix}+
\begin{bmatrix}
-2\rho_hV & 2\rho_vV\\
0 & -2\rho_{tw}V\\
2\rho_{tw}V & 0
\end{bmatrix}
\begin{bmatrix}
\dot{r}_y^{t,b} \\ \dot{r}_z^{t,b}
\end{bmatrix}+ \\
+\begin{bmatrix}
V^2(\rho_{tw}\rho_v-\rho'_h)-\rho_h\dot{V}  & \rho_v\dot{V}+\rho_{tw}\rho_hV^2 \\
-V^2(\rho_{tw}^2+\rho_h^2) & \rho_v\rho_hV^2-\rho_{tw}\dot{V} \\
\rho_{tw}\dot{V}+\rho_v\rho_hV^2 & -V^2(\rho_{tw}^2+\rho_h^2)
\end{bmatrix}
\begin{bmatrix}
r_y^{t,b} \\ r_z^{t,b}
\end{bmatrix} = \\
=\begin{bmatrix}
a_x^{imu}+a_z^{imu}\theta^{t,b}-a_y^{imu}\psi^{t,b}+g\theta^t-\dot{V} \\
a_y^{imu} + a_x^{imu}{\psi ^{t,b}} - a_z^{imu}{\varphi ^{t,b}} - g{\varphi ^t} - {\rho _h}{V^2}\\
a_z^{imu} - a_x^{imu}{\theta ^{t,b}} + a_y^{imu}{\varphi ^{t,b}} - g + {\rho _v}{V^2}
\end{bmatrix}
\end{aligned}
\end{equation}
where $V$ is the forward velocity of the body and $\rho _{h}$, $\rho _{v}$, and $\rho _{tw}$ are the horizontal, vertical and twist curvatures, respectively, of the track centerline. 

A further simplification that is also obtained in \cite{escalona2021} relates the "corrected" accelerometer signals with the absolute Euler angles of the body $b$, as follows:

\begin{equation} \label{Eq7_9}
{\bf{a}}^{imu}_{corr} = {\bf{a}}^{imu}_{filt} 
- \left[ {\begin{array}{*{20}{c}}
{\dot V}\\
{{\rho _h}{V^2}}\\
{ - {\rho _v}{V^2}}
\end{array}} \right]  \approx \it{g}\left[ {\begin{array}{*{20}{c}}
{ - \theta^{b}  }\\
\varphi^{b}  \\
1
\end{array}} \right]
\end{equation}

where ${\bf{a}}^{imu}_{corr}$ includes the "corrected" accelerometer signals. The correction is made to find a set of signals that are only due to the effect of gravity, eliminating to some extent the effect of the body acceleration. This correction is the result of low-pass filtering the signals to find ${\bf{a}}^{imu}_{filt}$ and subtracting and approximate value of the acceleration. See \cite{escalona2021} for more details. 

%% file: C_KalmanFilters.tex
\section{Kalman filters} \label{sec:KalmanFilters}

Two different algorithms have been developed to obtain the trajectory and attitude of the moving body from the IMU measurements.
\begin{itemize}
    \item[\tiny\textbullet] Two consecutive Kalman filters. The first one obtains the absolute orientation of the body $b$, while the second one provides its position relative to the TF.
    \item[\tiny\textbullet] One Kalman filter that includes the coupled \textit{orientation + position} kinematic model.
\end{itemize}
Please note that the term \textit{Kalman filter} is used in a broad sense throughout the paper, since the procedure applied to obtain the position and orientation of the moving body is actually a \textit{Kalman smoother}, which yields more accurate results~\cite{hartikainen}.

The chosen nomenclature for the \textit{state space models} is as follows:

\begin{equation}
\label{Eq_K2}
\left\{
\begin{aligned}
\textbf{x}_{k+1}=\textbf{Fx}_k+\textbf{q}_k
\\
\textbf{z}_k=\textbf{H}_k\textbf{x}_k+\textbf{r}_k
\end{aligned}
\right\},\quad \textrm{with} \quad
\left\{
\begin{aligned}
\textbf{q}_k\sim \textbf{N}(\textbf{0},\textbf{Q}) 
\\
\textbf{r}_k\sim \textbf{N}(\textbf{0},\textbf{R})
\end{aligned}
\right\}.
 \end{equation}
 
 \begin{itemize}
     \item[\tiny\textbullet] $\textbf{x}_k$: state vector of the system at time instant $k$.
     \item[\tiny\textbullet] $\textbf{z}_k$: measurements vector at time instant $k$.
     \item[\tiny\textbullet] $\textbf{F}$: transition matrix of the system.
     \item[\tiny\textbullet] $\textbf{H}_k$: measurements model matrix at time instant $k$.
     \item[\tiny\textbullet] $\textbf{Q}$: Process noise covariance matrix.
     \item[\tiny\textbullet] $\textbf{R}$: Measurements noise covariance matrix.
     \item[\tiny\textbullet] $\textbf{N}(\textbf{M},\textbf{S})$: Multivariate normal random variable with mean $\textbf{M}$ and covariance matrix $\textbf{S}$.
 \end{itemize}
 In the next subsections, all vectors and matrices in Eq.~\eqref{Eq_K2} will be specified for the different proposed algorithms, based upon the kinematic models described in Section~\ref{sec:KinematicModels}. Once these vectors and matrices are defined, the equations required to implement a Kalman filter or Kalman smoother algorithm can be found in any text on the subject, such as~\cite{hartikainen,sarkka}.

\subsection{Two consecutive Kalman filters}
\label{sec:Kalman_2consec}

\subsubsection{Kalman filter for the orientation}
\label{sec:Kalman_Or}

This filter estimates the time history of angles $\left\{\varphi^b, \theta^b, \psi^b\right\}$. It uses the kinematic model defined by equations~\eqref{Eq7_3} and~\eqref{Eq7_9}.

It should be noted that the yaw angle $\psi^b$ does not directly appear in ~\eqref{Eq7_3},~\eqref{Eq7_9}, but only its time derivative $\dot{\psi}^b$. This means that the numerical integration that allows obtaining $\psi^b$ from $\dot{\psi}^b$, which is part of the Kalman filter algorithm, may produce a drift effect by which, after certain time, the values of $\psi^b$ are totally erroneous. To solve this problem, it is helpful to include in the Kalman filter the additional equation
\begin{equation}
\label{Eq_K1}
    \psi^b=\psi^t,
\end{equation}
which represents a fictitious measurement. It might seem that the introduction of Eq.~\eqref{Eq_K1} will yield incorrect results, since it imposes the absolute yaw angle of the body to be exactly equal to that of the track. However, it should be kept in mind that all measurement equations have an associated noise--see $\textbf{r}_k$ in \eqref{Eq_K2}. As a consequence, the actual effect of Eq.~\eqref{Eq_K1} will be to keep $\psi^b$ close to $\psi^t$.  

The state vector for this filter is defined as 
\begin{equation}
\label{Eq_K3}
    \textbf{x}=
    \left[
    \begin{array}{*{20}{c}}
        \varphi^b & \dot{\varphi}^b & \theta^b & \dot{\theta}^b & \psi^b & \dot{\psi}^b  
    \end{array}
    \right]^T.
\end{equation}
Using ~\eqref{Eq7_3}, \eqref{Eq7_9}, \eqref{Eq_K1} and \eqref{Eq_K3}, all matrices required to build the state space model~\eqref{Eq_K2} can be obtained. It is convenient to make some comments about these matrices, which are shown in Appendix A:
\begin{itemize}
    \item [\tiny\textbullet] $\mathrm{\Delta}t$ represents the time interval between measurements.
   \item [\tiny\textbullet] The diagonal form of matrix $\textbf{R}$ corresponds to the assumption that the noises of the different measurements are statistically independent.
    \item [\tiny\textbullet] The form of matrix $\textbf{Q}$ represents the statistical assumption that variables $\dot{\varphi}^b(t)$, $\dot{\theta}^b(t)$ and $\dot{\psi}^b(t)$ evolve, between each pair of instants $k$ and $k+1$, following  Wiener processes. This is usually referred to in the literature as a CWNA model (\textit{Continuous White Noise Acceleration})~\cite{shalom,hartikainen}.
    \item [\tiny\textbullet] In order to define matrices $\textbf{Q}$ and $\textbf{R}$, the values of 7 parameters need to be specified. We collect these parameters in a vector $\textbf{p}$:
    \begin{equation}
    \label{Eq_K8}
        \textbf{p}=
        \left[
        \begin{array}{*{20}{c}}
        q_{\varphi} & q_{\theta} & q_{\psi} & R_{\omega} & R_{ax} & R_{ay} & R_{\psi}      
        \end{array}
        \right]^T.
    \end{equation}
\end{itemize}

\subsubsection{Kalman filter for the trajectory}

This filter estimates the time history of the body position with respect to the track, given by variables $\left\{r_y^{t,b}, r_z^{t,b}\right\}$. It uses the kinematic model defined by the last two equations in Eq.~\eqref{Eq9_3}. 

It is assumed here that the orientation of the body along time has already been estimated (see Section~\ref{sec:Kalman_Or}). Note that the Kalman filter for the orientation provides the absolute angles $\left\{\varphi^b, \theta^b, \psi^b\right\}$, while the relative angles $\left\{\varphi^{t,b}, \theta^{t,b}, \psi^{t,b}\right\}$ are actually more convenient to use in the Kalman filter for the trajectory (see Eq.~\eqref{Eq9_3}). The transformation from one set of angles to the other can be easily performed thanks to the following relations:
\begin{equation}
\label{Eq_K4}
    \begin{aligned}
    \varphi^{t,b}\approx \varphi^b-\varphi^t \\
    \theta^{t,b}\approx\theta^b-\theta^t \\
    \psi^{t,b}\approx\psi^b-\psi^t
    \end{aligned}
\end{equation}
where the assumption that $\left\{\varphi^{t,b}, \theta^{t,b}, \psi^{t,b}\right\}$ are small angles has been used.

As done in the Kalman filter for the orientation--see Eq.~\eqref{Eq_K1}--virtual sensors are introduced to avoid drift:
\begin{equation}
\label{Eq_K5}
    \begin{aligned}
        r_y^{t,b}=0 \\
        r_z^{t,b}=\delta
    \end{aligned}
\end{equation}
where $\delta$ represents the vertical distance at rest between the IMU and the track design centerline.

The state vector is defined in this case as

\begin{equation}
\label{Eq_K6}
    \textbf{x}=
    \left[
    \begin{array}{*{20}{c}}
        r_y^{t,b} & \dot{r}_y^{t,b} & \ddot{r}_y^{t,b} & r_z^{t,b} & \dot{r}_z^{t,b} & \ddot{r}_z^{t,b}
    \end{array}
    \right]^T.
\end{equation}

Using \eqref{Eq9_3}, \eqref{Eq_K4}, \eqref{Eq_K5} and \eqref{Eq_K6}, all matrices required to define the state space model~\eqref{Eq_K2} can be obtained. Let us make some remarks about these matrices, which are presented in Appendix B:
\begin{itemize}
   \item [\tiny\textbullet] The diagonal form of matrix $\textbf{R}$ represents the assumption that the noises of the different measurements are statistically independent.
    \item [\tiny\textbullet] The form of matrix $\textbf{Q}$ corresponds to the statistical assumption that variables $\ddot{r}_y^{t,b}$ and $\ddot{r}_z^{t,b}$ follow  Wiener processes between each pair of instants $k$ and $k+1$. This is usually known in the literature as a CWPA model (\textit{Continuous Wiener Process Acceleration})~\cite{shalom,hartikainen}.
    \item [\tiny\textbullet] The 6 scalar parameters needed to define matrices $\textbf{Q}$ and $\textbf{R}$ are collected in a vector $\textbf{p}$:
    \begin{equation}
       \label{Eq_K9}
        \textbf{p}=
        \left[
        \begin{array}{*{20}{c}}
        q_{y} & q_{z} & R_{y1} & R_{z1} & R_{y2} & R_{z2}   
        \end{array}
        \right]^T.
    \end{equation}
\end{itemize}

\subsection{One Kalman filter}
\label{sec:FullKalman}

This filter simultaneously estimates the absolute orientation of the moving body and its position relative to the track, given by variables $\{\varphi^b, \theta^b, \psi^b, r_y^{t,b}, r_z^{t,b}\}$. The algorithm exploits the kinematic model given by \eqref{Eq7_3}, \eqref{Eq9_3} and \eqref{Eq_K4}. Virtual measurements \eqref{Eq_K5} are also used here to avoid drift problems.

The state vector is defined as 
\begin{equation}
\label{Eq_K7}
    \textbf{x}=
    \left[
    \begin{array}{*{20}{c}}
        \varphi^b & \dot{\varphi}^b & \theta^b & \dot{\theta}^b & \psi^b & \dot{\psi}^b & r_y^{t,b} & \dot{r}_y^{t,b} & \ddot{r}_y^{t,b} & r_z^{t,b} & \dot{r}_z^{t,b} & \ddot{r}_z^{t,b}
    \end{array}
    \right]^T.
\end{equation}

Using  \eqref{Eq7_3}, \eqref{Eq9_3}, \eqref{Eq_K4}, \eqref{Eq_K5} and \eqref{Eq_K7}, all matrices required to formulate the state space model~\eqref{Eq_K2} can be obtained. It is pertinent to make some clarifications about these matrices, which are shown in Appendix C:
\begin{itemize}
   \item [\tiny\textbullet] The diagonal form of matrix $\textbf{R}$ represents the assumption of statistically independent noises for the different measurements.
    \item [\tiny\textbullet] The form of matrix $\textbf{Q}$ corresponds to the statistical assumption that variables $\dot{\varphi}^b(t)$, $\dot{\theta}^b(t), \dot{\psi}^b(t), \ddot{r}_y^{t,b}$ and $\ddot{r}_z^{t,b}$ evolve, between each pair of instants $k$ and $k+1$, following  Wiener processes.
    \item [\tiny\textbullet] Eleven parameters are required to fully define covariance matrices $\textbf{Q}$ and $\textbf{R}$ . We gather these parameters in a vector $\textbf{p}$:
    \begin{equation}
       \label{Eq_K10}
        \textbf{p}=
        \left[
        \begin{array}{*{20}{c}}
        q_{\varphi} & q_{\theta} & q_{\psi} & q_y & q_z & R_{\omega} & R_{x} & R_{y1} & R_{z1} & R_{y2} & R_{z2}      
        \end{array}
        \right]^T.
    \end{equation}
\end{itemize}
Before concluding Section~\ref{sec:KalmanFilters}, it is relevant to note that we also tested augmented versions of the 3 presented Kalman filters that incorporate additional state variables to model sensor biases. The corresponding equations have not been included here because the results were not satisfactory, as discussed in Section~\ref{sec:discussion}.

%% file: D_EstimationCovariances.tex
\section{Methods for the estimation of covariance matrices} \label{sec:EstimationCovariance}

In the last section, two different Kalman filtering approaches have been presented with the aim of estimating the trajectory and attitude of a generic body $b$ moving along a railway track. To be able to use these algorithms, the first step is to fully define covariance matrices $\textbf{Q}$ and $\textbf{R}$ in Eq.\eqref{Eq_K2}. 

We use $\textbf{p}=[\begin{array}{*{20}{c}}
        p_1 & p_2 & ... & p_M 
    \end{array}]^T$ to denote a vector containing the scalar parameters that determine matrices $\textbf{Q}$ and $\textbf{R}$. The number of elements $M$ in $\textbf{p}$ depends on the specific Kalman Filter under consideration--see Eqs. \eqref{Eq_K8}, \eqref{Eq_K9} and \eqref{Eq_K10}.

Before proceeding to describe the proposed strategies to estimate vector $\textbf{p}$, some clarifications are appropriate. The developments presented in Sections \ref{sec:KinematicModels} and \ref{sec:KalmanFilters} are completely general, in the sense that they can be used in any application that requires estimating the position and orientation of a vehicle moving along a railway track. On the other hand, these techniques have been applied in the current research to the specific problem of measuring track irregularities. A \textit{Track Geometry Measuring System} (TGMS) has been developed by the authors, combining computer vision with the estimation of the trajectory and attitude of the vehicle using inertial sensors. The details of this TGMS technology can be found in~\cite{escalona2021} and are summarized in Section~\ref{sec:Experiments}. The present section expounds two alternative procedures to estimate vector $\textbf{p}$. The first one is specific to the TGMS application, while the second one is as general as Sections \ref{sec:KinematicModels} and \ref{sec:KalmanFilters}:
\begin{itemize}
    \item [\tiny\textbullet] Known-output method. The covariance matrices are tuned through a comparison between the track irregularities obtained with the TGMS and a set of \textit{reference irregularities}.
    \item [\tiny\textbullet] Constrained maximum likelihood estimation. This technique uses the IMU measurements to obtain the \textit{most likely} vector $\textbf{p}$.
\end{itemize}
In both cases, a global optimization of certain objective function is conducted, with the components of vector $\textbf{p}$ varying within a specified range. In order to choose appropriate range limits, consider first that every parameter $p_i$ must be positive due to their physical meaning--see Section~\ref{sec:KalmanFilters}. On the other hand, the range should be wide enough to encompass a meaningful optimum of the objective function. After several trials, the following range has been found to produce accurate results:
\begin{equation}
\label{Eq_C3}
    10^{-4}\leq p_i \leq 10^4 \qquad \textrm{for} \ \ i=1,...,M
\end{equation}
with each $p_i$ measured in SI.

After testing different global optimization techniques included in the \textit{MATLAB Optimization Toolbox}, the method that was found to work best for this application was the \textit{Surrogate Optimization}~\cite{queipo}. This scheme does not require an initial point for the optimum search, but only the ranges where the parameters can vary and, optionally, one or more conditions that they must fulfill.

\subsection{Known-output method}
\label{sec:known-output}

In order to apply the \textit{Known-Output Method} (KOM), the vehicle with the operating TGMS has to travel along a track whose irregularities have already been obtained through some other reliable methodology (reference irregularities). For the present work, these reference irregularities were obtained by using an LVDT, an inclinometer and a Total Station, as described in detail in~\cite{urda2020}. 

It is convenient at this point to recall that we are interested in the 4 irregularities that are usually controlled in the railway industry: 
\begin{itemize}
    \item [\tiny\textbullet] Alignment and Vertical Profile (absolute irregularities)
    \item [\tiny\textbullet] Gauge Variation and Cross Level (relative irregularities)
\end{itemize}
See~\cite{escalona2021} for precise definitions of these terms.

The optimization algorithm searches for the vector $\textbf{p}$ that minimizes the RMS of the difference between the TGMS irregularities and the reference irregularities. Thus, the objective function ($\textrm{OF}$) for the optimization is defined as
\begin{equation}
\label{Eq_C1}
\begin{aligned}
    \textrm{OF}=\textrm{RMS}[\textrm{TGMS Alignment}(s)-\textrm{Reference Alignment}(s)]+\\
    \textrm{RMS}[\textrm{TGMS Vertical Profile}(s)-\textrm{Reference Vertical Profile}(s)]+\\
    \textrm{RMS}[\textrm{TGMS Cross Level}(s)-\textrm{Reference Cross Level}(s)].
\end{aligned}
\end{equation}
Note that the gauge variation is not included in \eqref{Eq_C1}. The reason is that this irregularity has been found to depend almost exclusively on the computer vision of the TGMS, with a negligible influence of the vehicle trajectory and attitude.

Let us make some remarks about the described strategy:
\begin{itemize}
    \item [\tiny\textbullet] From a pragmatic point of view, this approach has a clear disadvantage: when installing the TGMS on a new commercial vehicle, it would need to travel along a track whose irregularities were reliably known beforehand, which is not easy to find in practice. Furthermore, the appeal of the TGMS technology could get limited by the need of a complex initial calibration.
    \item [\tiny\textbullet] This procedure  relies on the assumption that, once the covariance matrices have been estimated for a certain track, with certain levels of irregularity and a vehicle circulating at certain speed, the Kalman filters will still produce accurate results for other tracks, with different irregularities and different vehicle speeds. The validity of this hypothesis is not evident. In fact, the results presented in Section~\ref{sec:Results} suggest that the mentioned assumption does not always hold. In other words, the covariance matrices obtained for a specific railway track could be ineffective to estimate the irregularities of other tracks. 
\end{itemize}

\subsection{Constrained maximum likelihood estimation}
\label{sec:MLE}

Maximum Likelihood Estimation (MLE) is a very common approach for tuning the parameters of statistical models--the sense of the word \textit{constrained} in the title will be seen at the end of the section. The basic idea, for those readers who are not familiar with this kind of methods, is the following. Consider a generic multivariate random variable $\textbf{y}$ whose probability density function, $p(\textbf{y}|\textbf{p})$, depends on a set of parameters $\textbf{p}$. Assume these parameters are unknown and we are interested in estimating them. According to the MLE strategy, the estimation includes 2 steps:
\begin{itemize}
    \item [\tiny\textbullet] Carry out the experiment that is modelled by the probability density function $p(\textbf{y}|\textbf{p})$. Let $\textbf{y}_0$ denote the obtained result (observation).
     \item [\tiny\textbullet] Estimate $\textbf{p}$ as the set of parameters that maximizes $p(\textbf{y}_0|\textbf{p})$. That is to say, we choose the parameters that are most likely to generate the observed data.
\end{itemize}
For a more rigorous and complete explanation of the MLE concept, see~\cite{miura}.

The generic scheme outlined above can be directly particularized to the trajectory and attitude estimation that is being investigated. While the vector of parameters $\textbf{p}$ is given in Eqs.~\eqref{Eq_K8}, \eqref{Eq_K9} and \eqref{Eq_K10}, the set of observed data $\textbf{y}_0$ represents the sequence of accelerations, angular velocities, forward speeds and positions along the track registered during a specific ride of the moving body $b$ (Fig.~\ref{fig:railKinematics3D}). Finally, the probability density of the observations $p(\textbf{y}|\textbf{p})$ can be obtained from the model equations \eqref{Eq_K2}, as shown in~\cite{abbeel}. 

In summary, given a set of observations $\textbf{y}_0$, the proposed global optimization algorithm will explore the space of parameters $\textbf{p}$ within the range specified in Eq.~\eqref{Eq_C3}, searching for those parameter values that maximize $p(\textbf{y}_0|\textbf{p})$. The specific equations that need to be appended to the Kalman Filter algorithm in order to calculate $p(\textbf{y}_0|\textbf{p})$ can be found in~\cite{abbeel}.

Some comments about this approach are pertinent:
\begin{itemize}
    \item [\tiny\textbullet] It is usual in the Kalman filtering literature to conduct the likelihood maximization through a specific procedure called \textit{Expectation Maximization algorithm} (EM algorithm) \cite{berg,EM}. This method has been implemented by the authors for the trajectory and attitude estimation, but the attained results have not been satisfactory. This is why the presented global optimization scheme has been chosen instead. In any case, it should be noted that the EM algorithm intends to find a local maximum of the likelihood function, which in principle makes a global maximization technique preferable.
    \item [\tiny\textbullet] When the described methodology is used for the TGMS application, it has a plain advantage over the one set forth in Section~\ref{sec:known-output}, since in this case there is no need of a previously known set of irregularities to tune the covariance matrices. Consider further the following benefit: assume that a railway vehicle, with an operating TGMS, performs two rides on two very different tracks, with different irregularity levels and moving at different forward speeds. As was commented in Section~\ref{sec:known-output}, it is not obvious that one only pair of matrices $\textbf{Q}$ and $\textbf{R}$ could provide accurate results for both tracks. The MLE would take this automatically into account by obtaining one set of \textit{most likely} covariance matrices for each of the rides.
\end{itemize}

\subsubsection{Constraints imposed for the optimization}

The experimental results obtained during this research have shown that the MLE, as has been described, can produce inaccurate outcomes in this application. That is to say, the most likely covariance matrices do not necessarily provide a good position and attitude estimation. This problem has been solved by slightly guiding the search for the optimum considering some conditions on the parameters. In particular, accurate results have been found--as will be shown in Section~\ref{sec:Results}--thanks to the following constraints:
\begin{itemize}
   \item[\tiny\textbullet] Two consecutive Kalman Filters
   \subitem -- Kalman Filter for the Orientation 
   \begin{equation}
       R_{ay}\leq q_{\varphi}/10
   \end{equation}
   \subitem -- Kalman Filter for the Trajectory
   \begin{equation}
       R_{y1}\leq R_{y2}/10
   \end{equation}
   \item[\tiny\textbullet] One Kalman Filter
    \begin{equation}
    \label{Eq_C2}
       R_{y1}\leq R_{y2}/10
   \end{equation}
\end{itemize}
with all parameters measured in SI.

The interpretation of the above inequalities can be readily seen by recalling the physical meaning of each of the parameters. For example, in the case of Eq.~\eqref{Eq_C2}, $R_{y1}$ represents the variance of the noise associated with the accelerometer signal in $y$ direction, while $R_{y2}$ is the variance of the noise associated with the virtual measurement $r_y^{t,b}=0$, as can be found in Section~\ref{sec:FullKalman}. Considering that the noise variance for each sensor represents the uncertainty of the corresponding measurement, the meaning of Eq.~\eqref{Eq_C2} could be stated as follows: \textit{measuring all parameters in SI, the Kalman filter must have a considerably higher confidence in the accelerometer measurements in $y$ direction than it has in the virtual measurements $r_y^{t,b}=0$}. This way, we avoid the optimization process from leading to a solution of the type $r_y^{t,b}(t)\approx0$ due to an excessive weight of the fictitious position measurements in the calculations.

It can be said after the above considerations that, instead of a plain MLE, the proposed approach falls within the category of \textit{Constrained Maximum Likelihood estimation} (CML estimation)~\cite{const1,const2}.

%% file: E_Experiments.tex
\section{TGMS method and experimental setup} \label{sec:Experiments}

\begin{figure}[htbp!]
	\centering
	\includegraphics[width=0.8\linewidth]{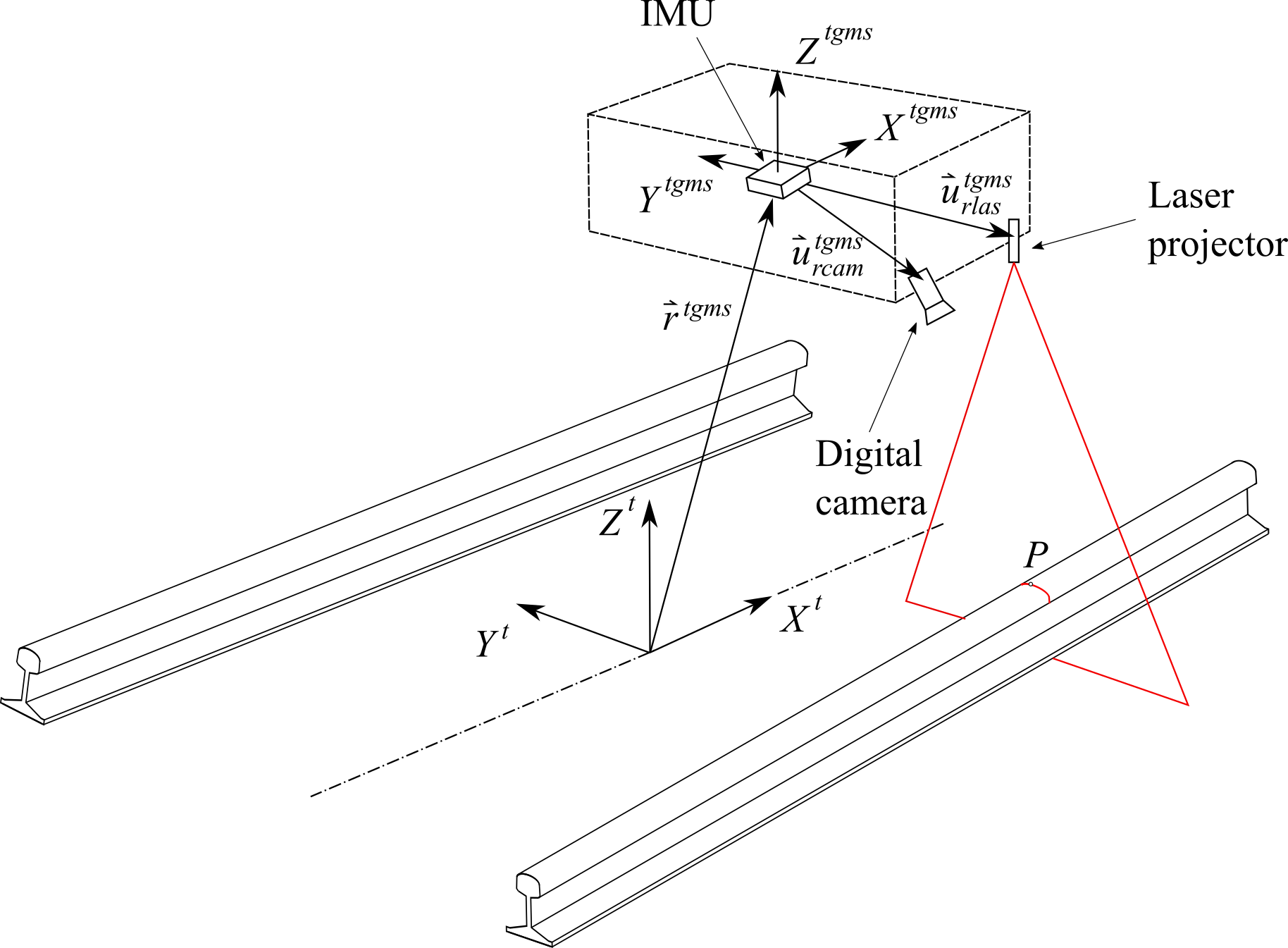}
	\caption{Track geometry measuring system}
	\label{fig:TGMSonTrack}
\end{figure}

\begin{figure}[htbp!]
	\centering
	\includegraphics[width=0.8\linewidth]{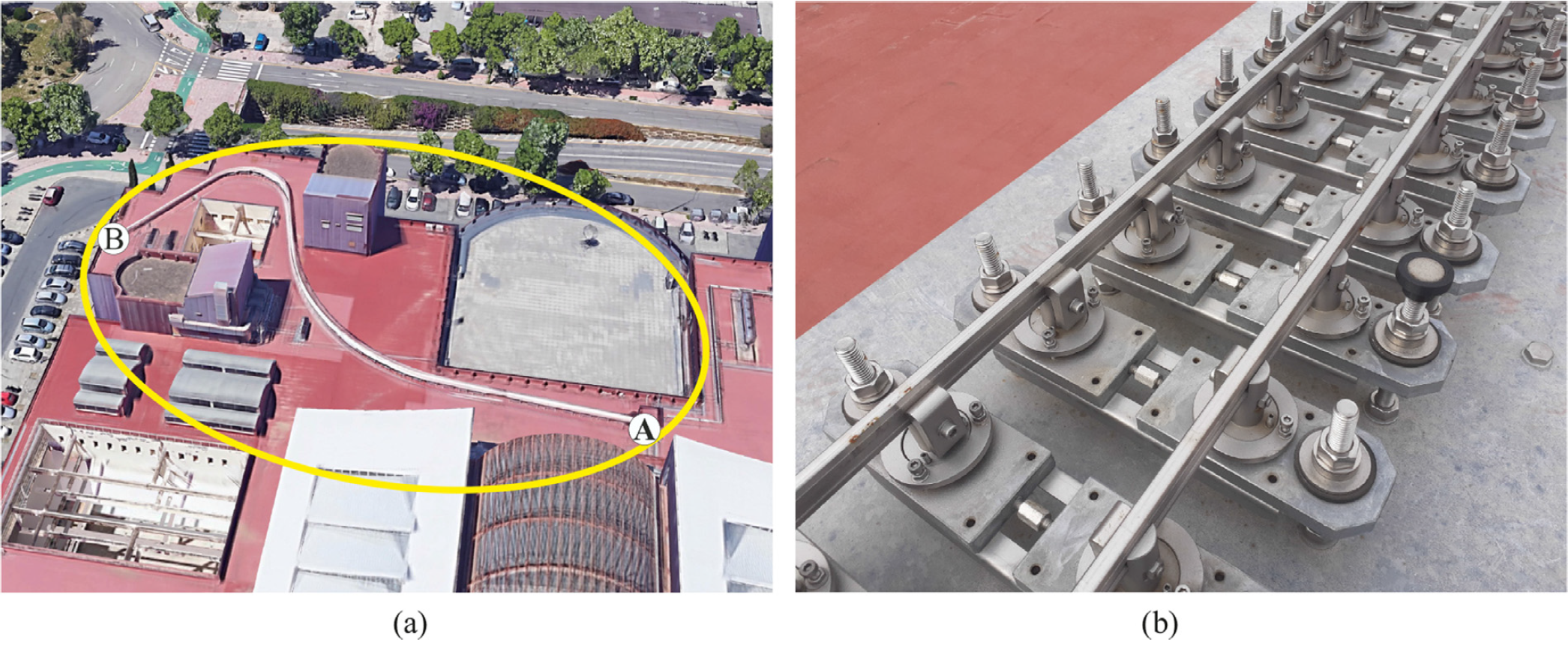}
	\caption{Scale track: (a) Aerial view, (b) detail of track supports}
	\label{fig:scaleTrack}
\end{figure}

\begin{figure}[htbp!]
	\centering
	\includegraphics[width=0.8\linewidth]{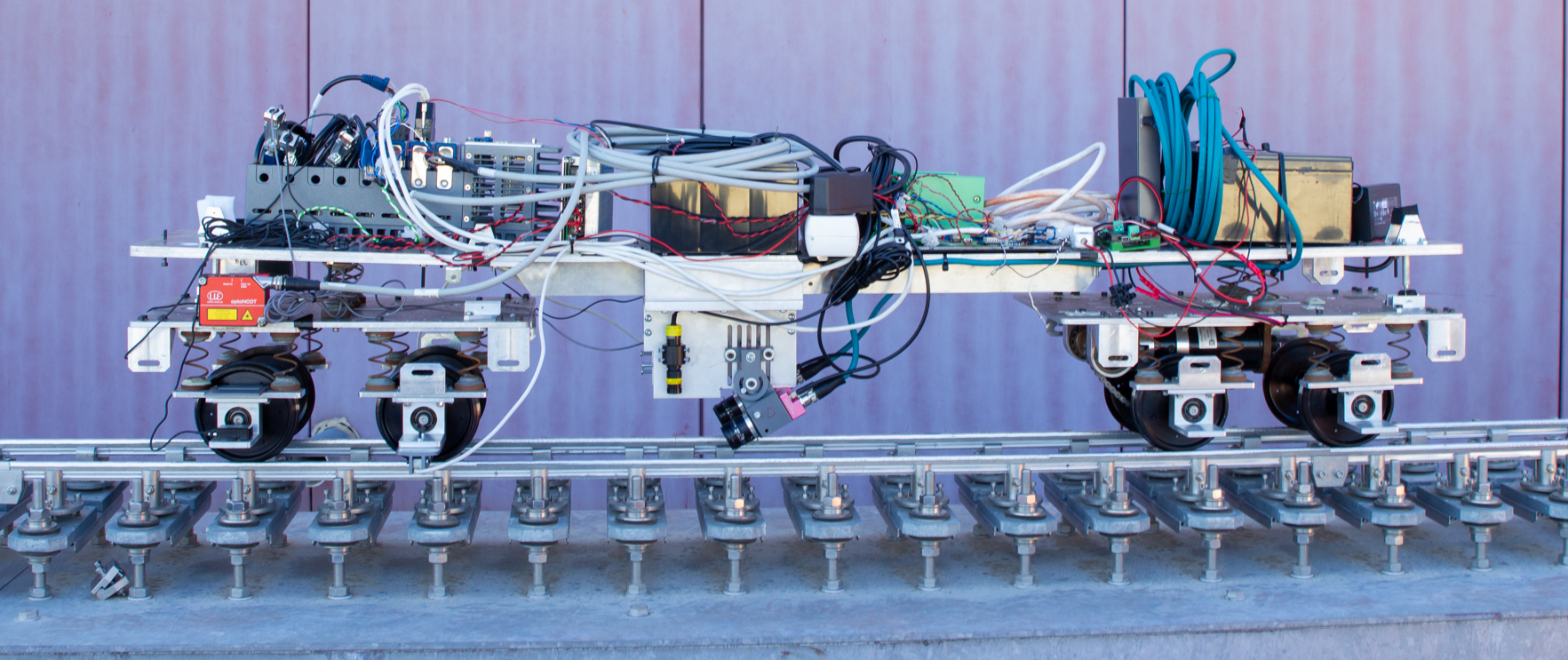}
	\caption{Scale vehicle with TGMS}
	\label{fig:vehiculoGrande}
\end{figure}

The TGMS is a technology developed by the authors for the measurement of the irregularities of rail tracks. It combines inertial sensors with computer vision. The sensors include an IMU, two video cameras, two laser projectors and an encoder for the odometry and calculation of the forward velocity. This equipment is schematically represented in Fig \ref{fig:TGMSonTrack}.  This technology is explained in detail in \cite{escalona2021}. Results shown in this paper are obtained with a 1:10 scale experimental facility installed at the School of Engineering of the University of Seville, Spain. Figure \ref{fig:scaleTrack} shows the 90 m-scale track that has been designed to create an arbitrary distribution of irregularities. To this end, the sleepers have been substituted with 4-dof mechanisms that allow the in-plane motion of both rail cross-sections. Figure \ref{fig:vehiculoGrande} shows the scale vehicle used in the experiments that incorporates the TGMS. The vehicle has a classical structure of 4 rigid wheelsets, two bogie frames and one carbody, with primary and secondary suspensions. In this figure, two video cameras and one of the laser projectors can be distinguished in the central part of the carbody. The IMU cannot be seen in the figure, but it is located also in the central part of the carbody. Therefore in this case, body $b$, whose trajectory and orientation will be estimated, is the vehicle carbody. It is important to emphasize that the TGMS allows the installation of the sensors in any body of the vehicle. \\

The algorithms developed in \cite{escalona2021} for the estimation of irregularities with the TGMS require the evaluation of the position and orientation of the IMU. While the vehicle trajectory has a strong influence on the estimation of alignment and vertical profile, the vehicle orientation (particularly the roll angle) is very relevant for the estimation of alignment and cross level--the effect of both trajectory and orientation on the gauge estimation is negligible, as commented in Section~\ref{sec:known-output}. Thanks to the improvements in the trajectory and attitude estimation described in this paper, the obtained alignment, vertical profile and cross level will be shown to be notably more accurate than those presented in~\cite{escalona2021}.

%% file: F_Results.tex
\section{Results} \label{sec:Results}

This section presents the experimental results that have been obtained when applying the procedures developed in this paper to the problem of track geometry measurement, as described in Section~\ref{sec:Experiments}. For the details on how the trajectory and attitude estimation is used within the TGMS technology to obtain track irregularities, see~\cite{escalona2021}.

\begin{figure}[htbp!]
  \centering
    \includegraphics[width=0.7\textwidth]{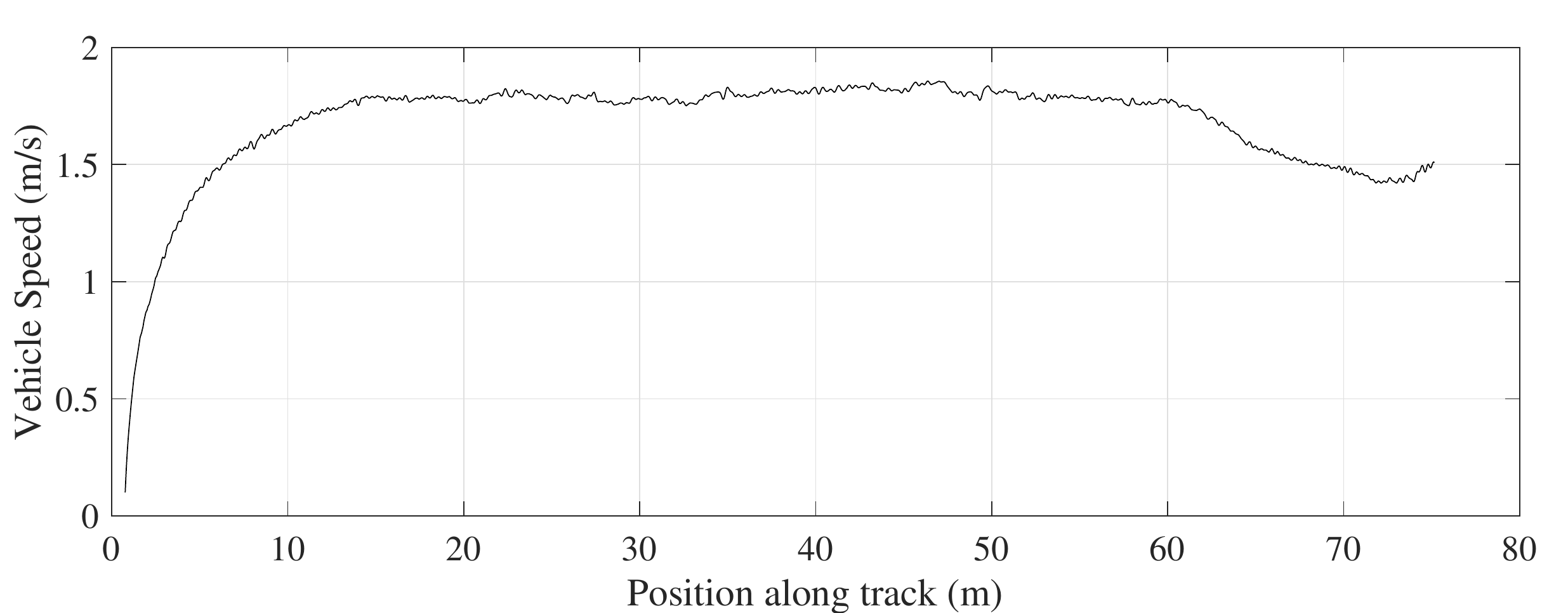}
    \caption {Forward velocity of the scale vehicle during the test}
  \label{fig:Results_Speed}
\end{figure}

All shown results correspond to one ride of the instrumented vehicle along the scale track, with the forward velocity shown in Fig.~\ref{fig:Results_Speed}.

Some of the irregularity profiles displayed throughout this section are filtered. Whether or not each particular irregularity has been filtered for the representation is specified in the corresponding figure caption (Figs.~\ref{fig:Results_GV_2F}-\ref{fig:Results_VP_Rob}). Filtered irregularities only preserve wavelengths between $\SI{0.3}{\metre}$ and $\SI{7}{\metre}$. This is in accordance with European Standard~\cite{EN-Norm}, which states that irregularity wavelengths between $\SI{3}{\metre}$ and $\SI{70}{\metre}$ are the ones directly associated with railway vehicles safety--note that the 1:10 scale of the track has been applied to the wavelength range.

\subsection{Two consecutive Kalman filters}

\begin{figure}[h!]
  \centering
    \includegraphics[width=0.7\textwidth]{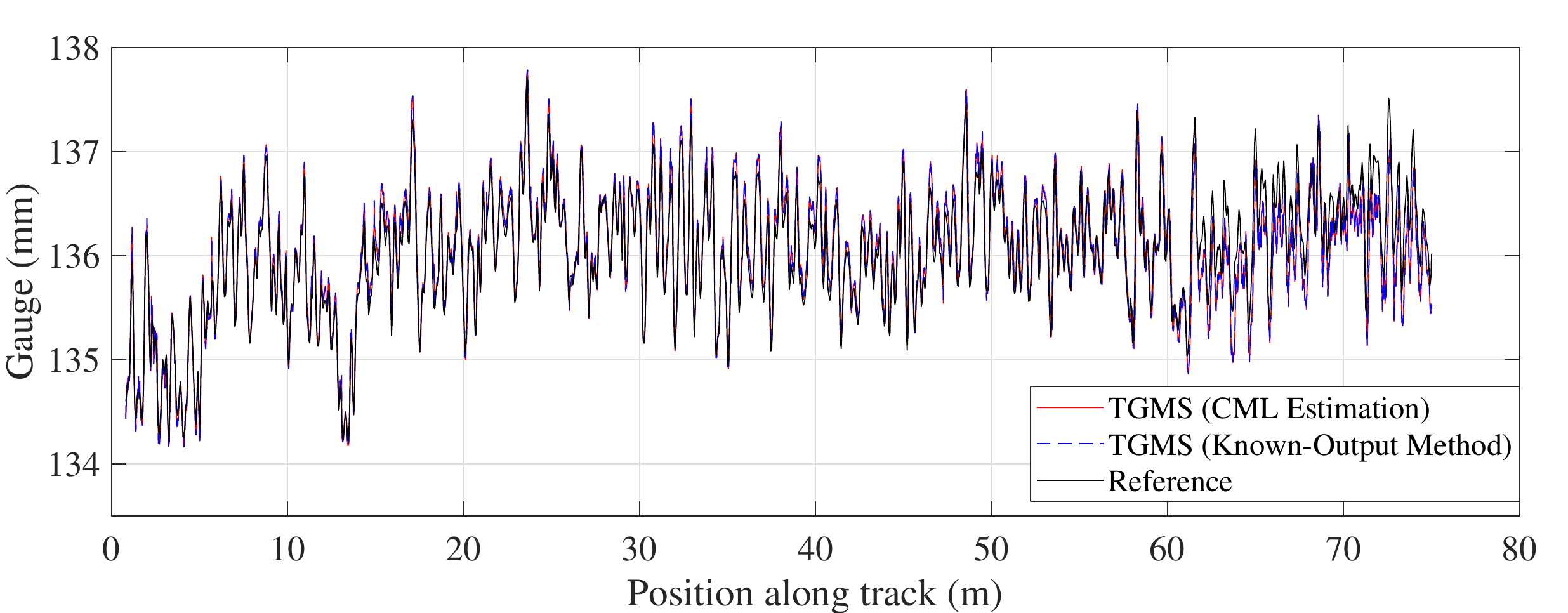}
    \includegraphics[width=0.7\textwidth]{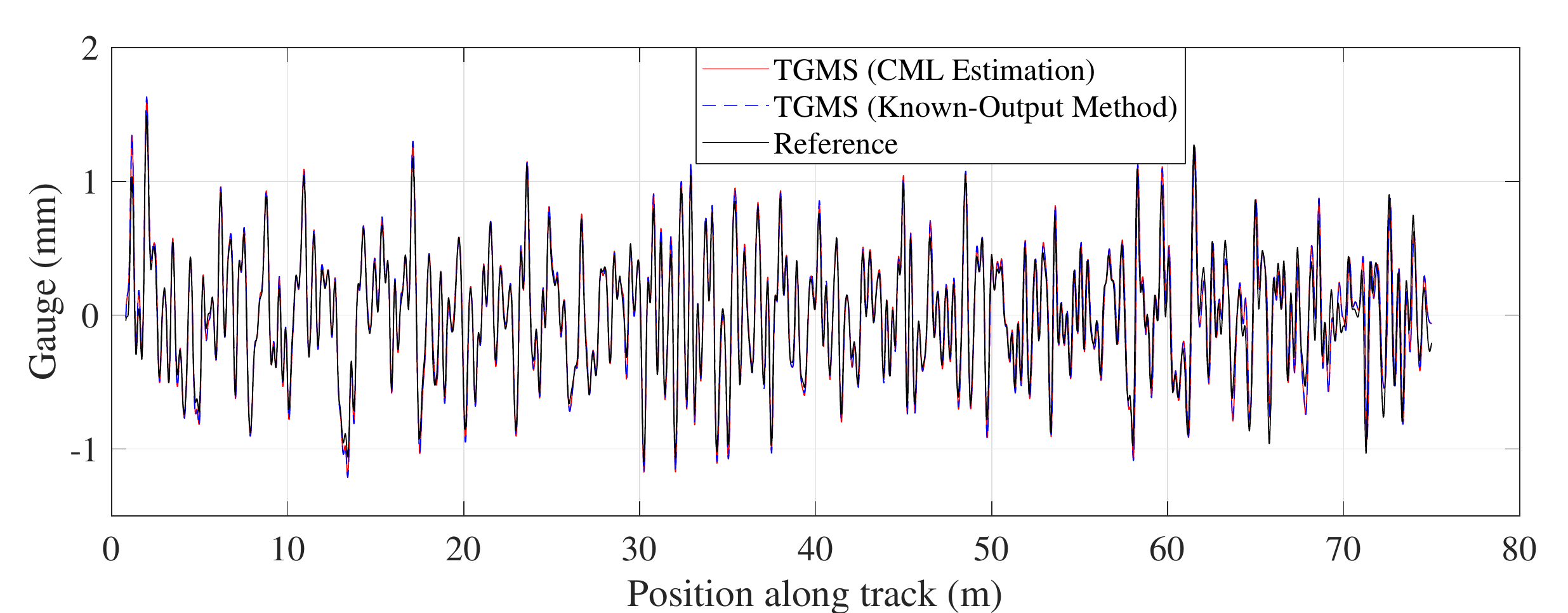}
  \caption {Track Gauge obtained with 2 consecutive Kalman filters, compared to the Reference Gauge. Top: Unfiltered. Bottom: Filtered.}
  \label{fig:Results_GV_2F}
\end{figure}

\begin{figure}[h!]
  \centering
    \includegraphics[width=0.7\textwidth]{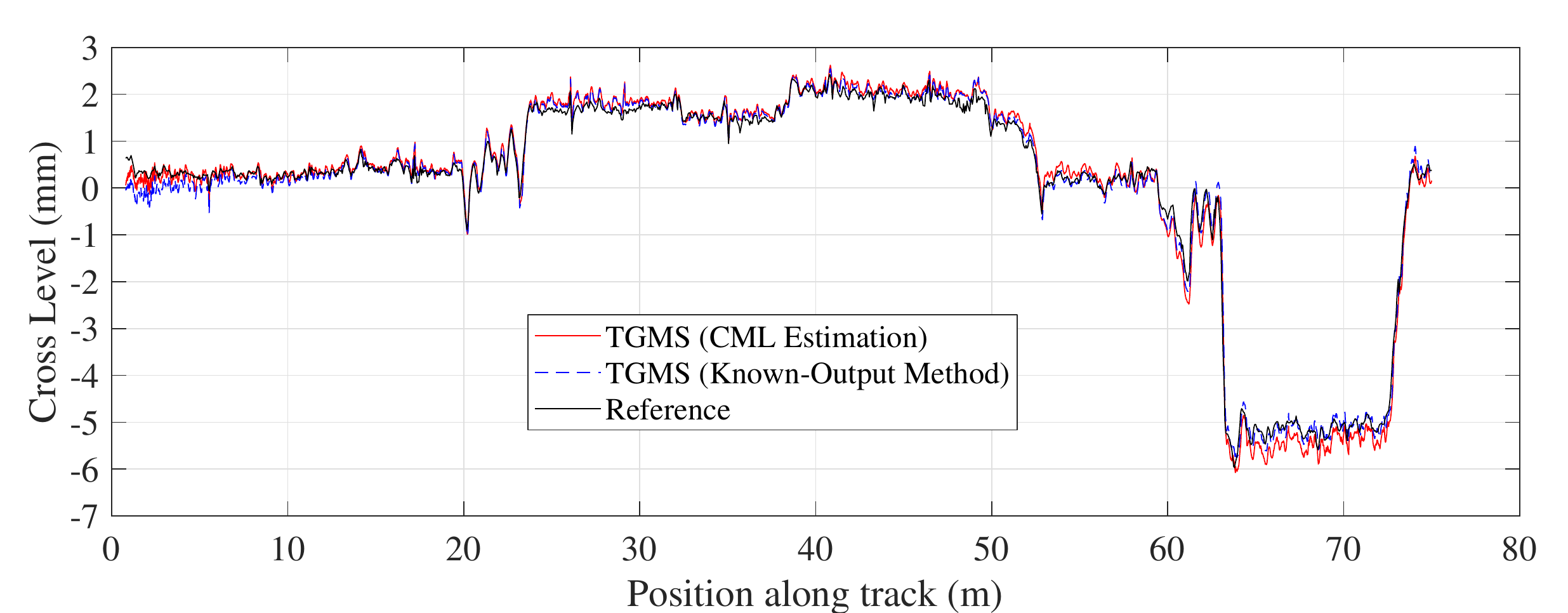}
    \includegraphics[width=0.7\textwidth]{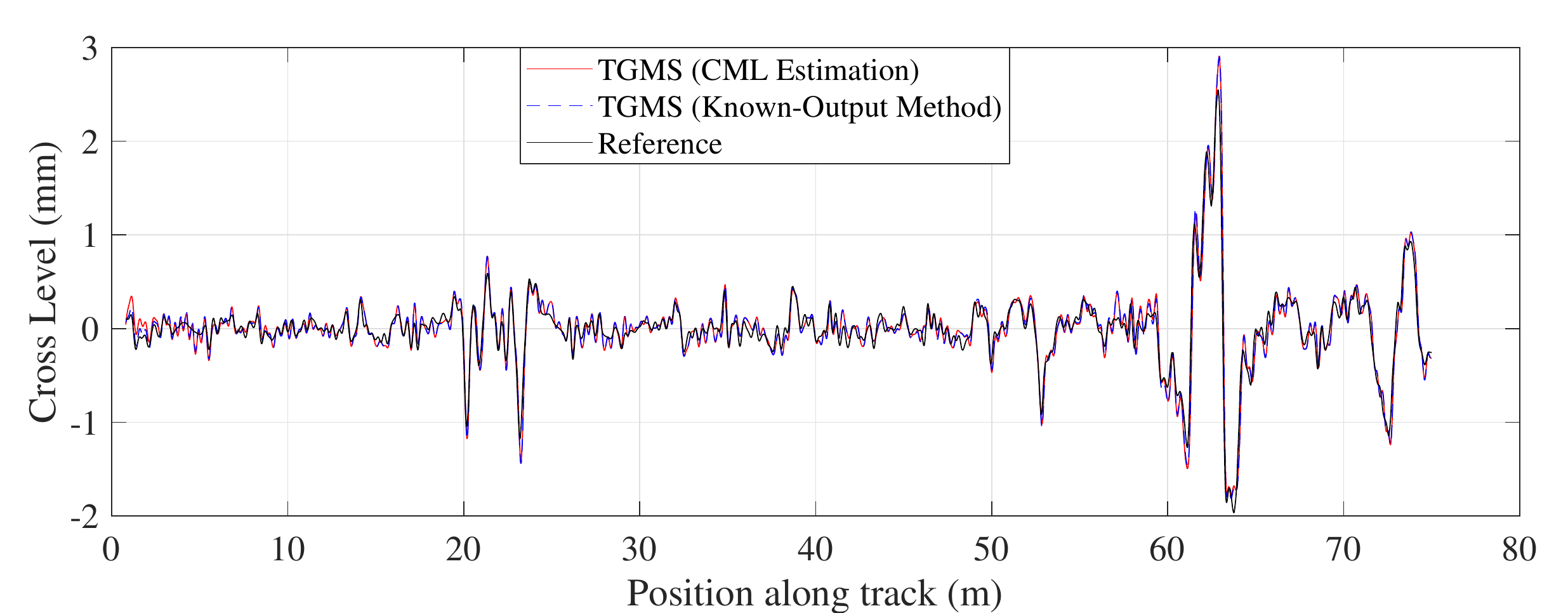}
  \caption{Cross Level obtained with 2 consecutive Kalman filters, compared to the Reference Cross Level. Top: Unfiltered. Bottom: Filtered.}
  \label{fig:Results_CL_2F}
\end{figure}

\begin{figure}[h!]
  \centering
    \includegraphics[width=0.7\textwidth]{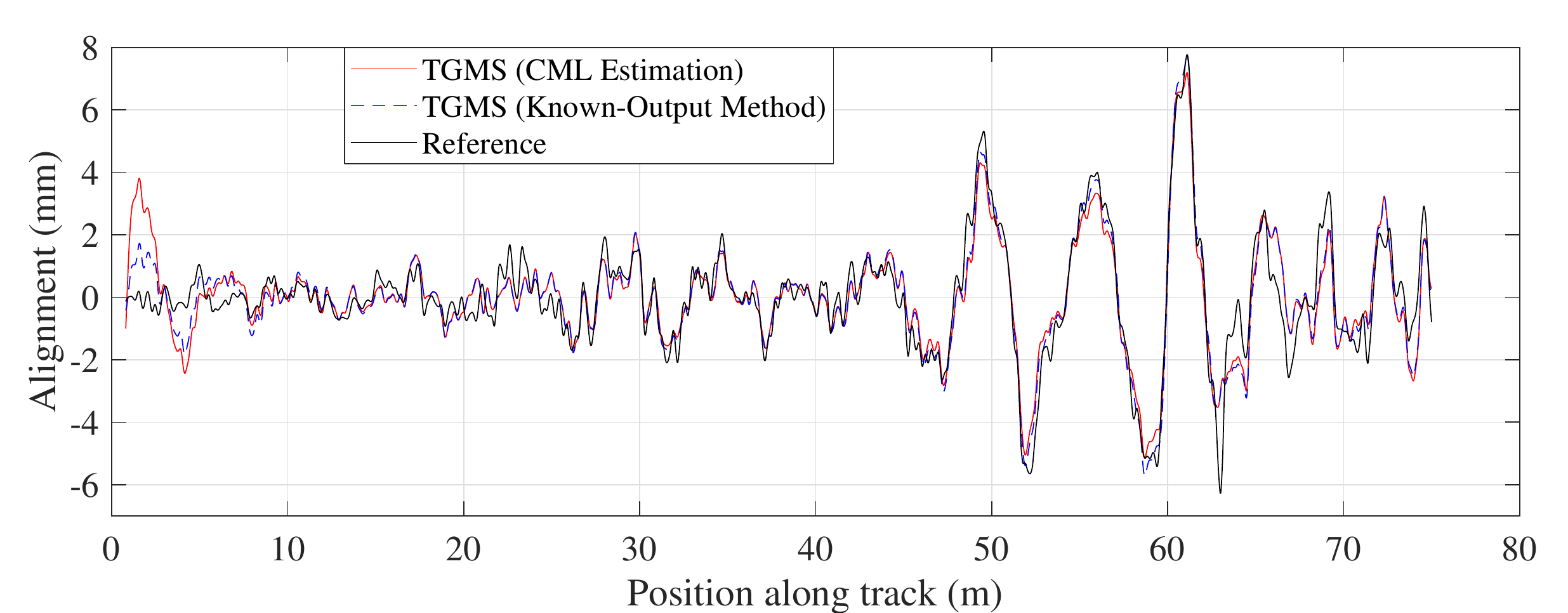}
  \caption{Alignment obtained with 2 consecutive Kalman filters, compared to the Reference Alignment (all signals are filtered).}
  \label{fig:Results_AL_2F}
\end{figure}

\begin{figure}[h!]
  \centering
    \includegraphics[width=0.7\textwidth]{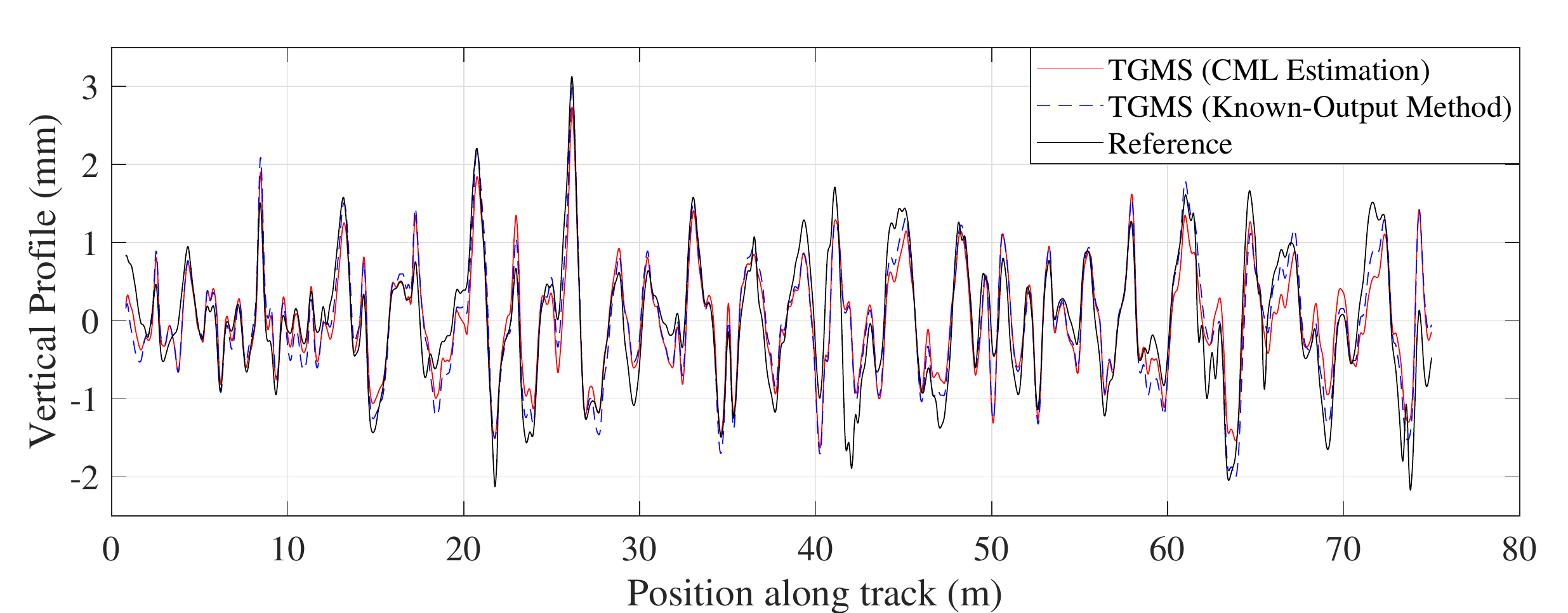}
  \caption{Vertical Profile obtained with 2 consecutive Kalman filters, compared to the Reference Vertical Profile (all signals are filtered).}
  \label{fig:Results_VP_2F}
\end{figure}

The irregularities shown in Figs.~\ref{fig:Results_GV_2F}-\ref{fig:Results_VP_2F} correspond to the application of 2 consecutive Kalman filters for the trajectory and attitude estimation, as presented in Section~\ref{sec:Kalman_2consec}. We compare the accuracy of the CML estimation for covariance matrices (Section~\ref{sec:MLE}) with that of the KOM (Section~\ref{sec:known-output}).

\subsection{One Kalman filter}

\begin{figure}[h!]
  \centering
    \includegraphics[width=0.7\textwidth]{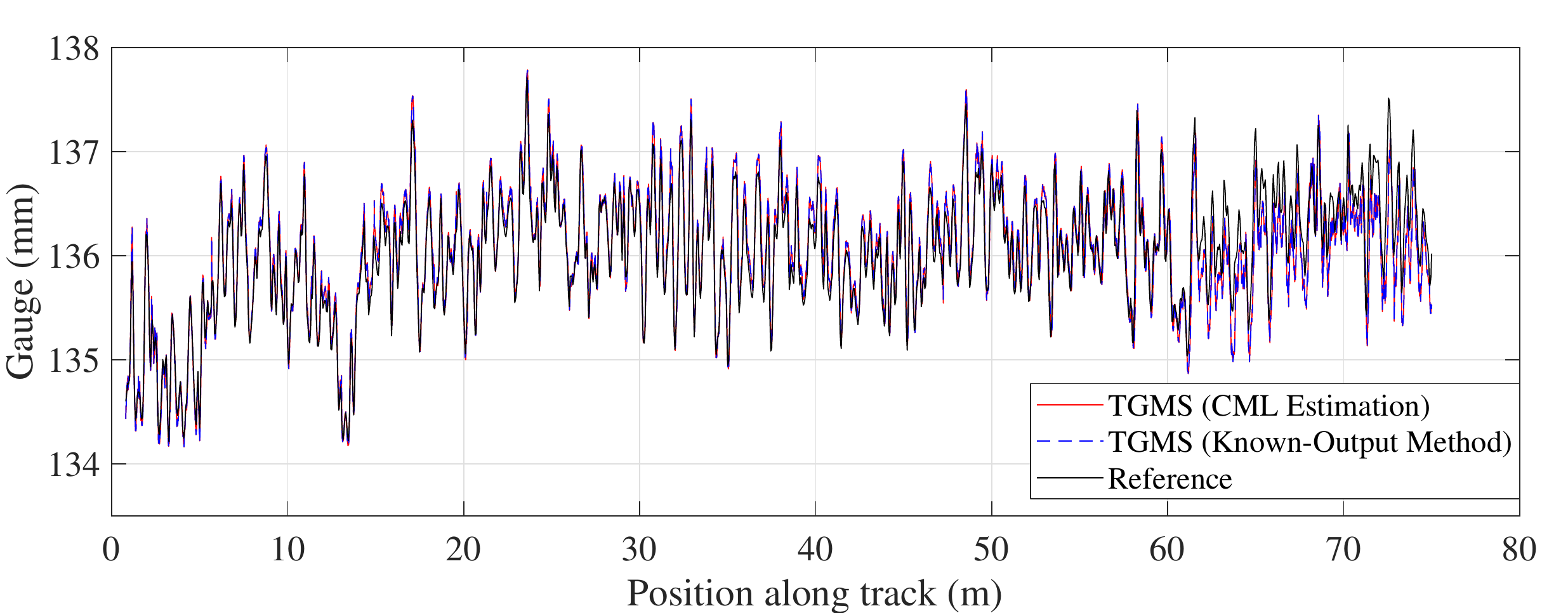}
    \includegraphics[width=0.7\textwidth]{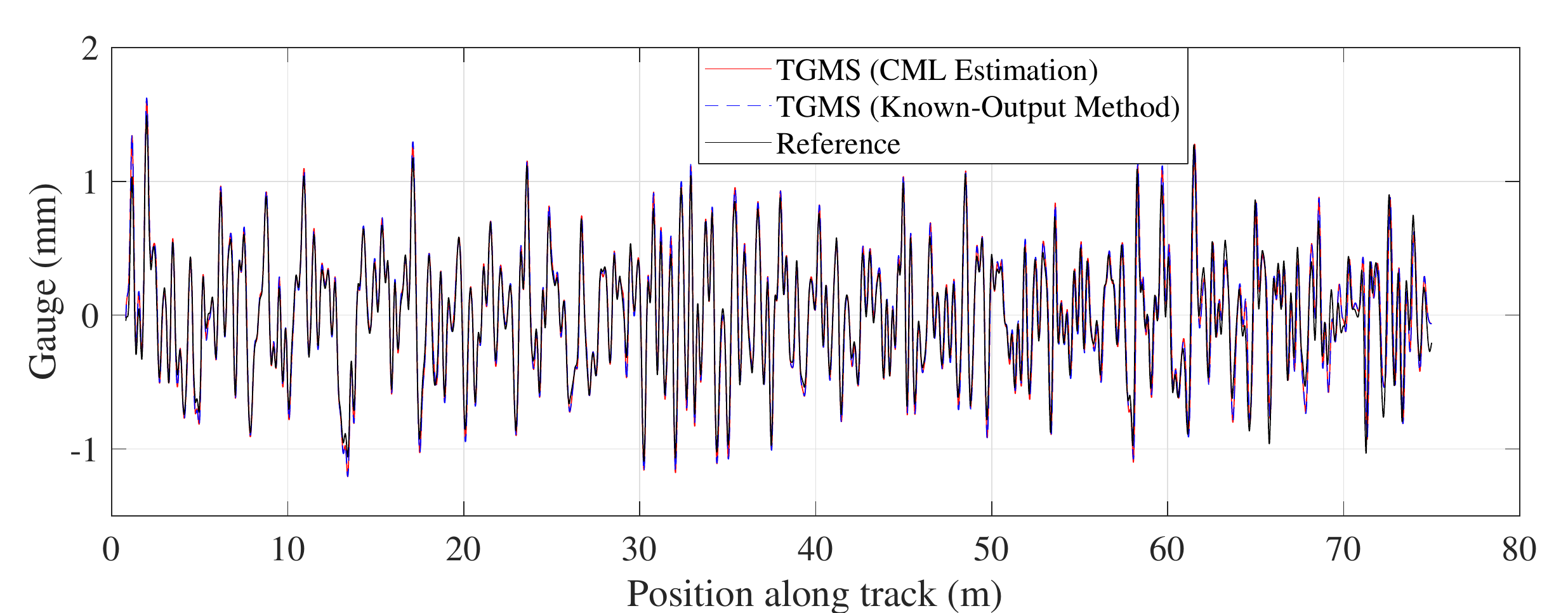}
  \caption {Track Gauge obtained with one Kalman filter, compared to the Reference Gauge. Top: Unfiltered. Bottom: Filtered.}
  \label{fig:Results_GV_1F}
\end{figure}

\begin{figure}[h!]
  \centering
    \includegraphics[width=0.7\textwidth]{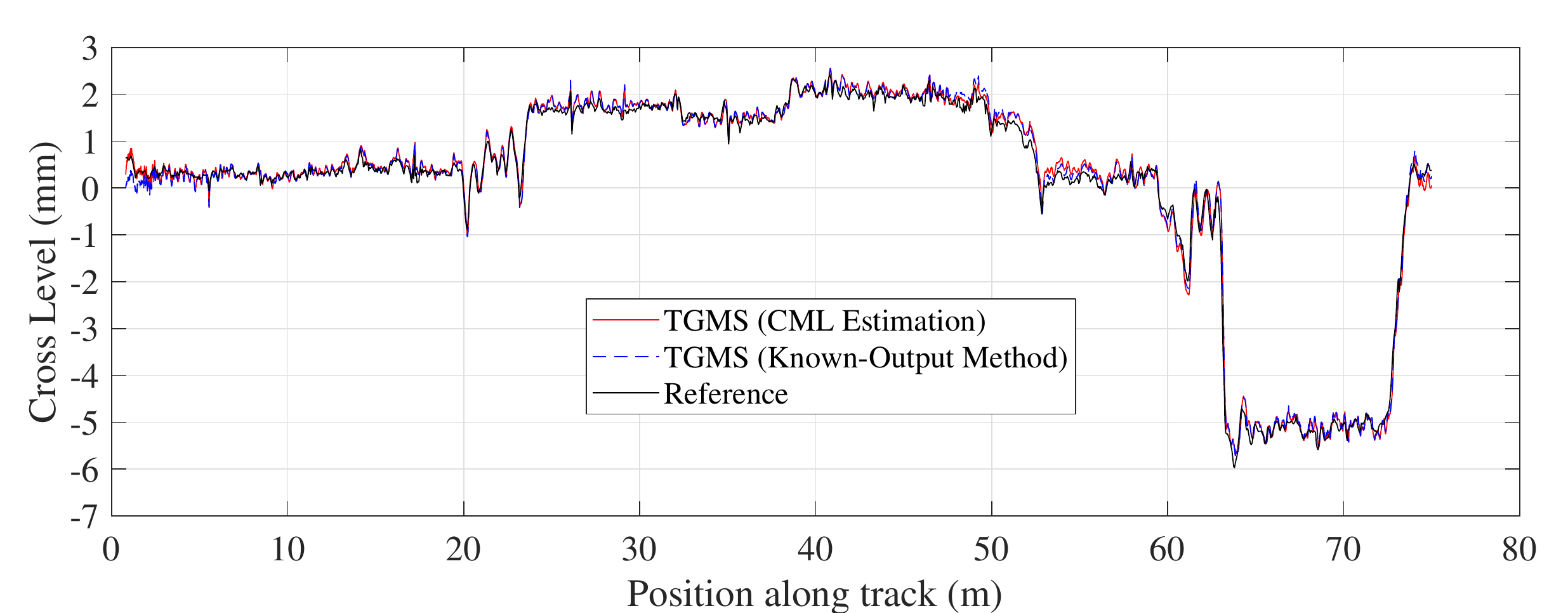}
    \includegraphics[width=0.7\textwidth]{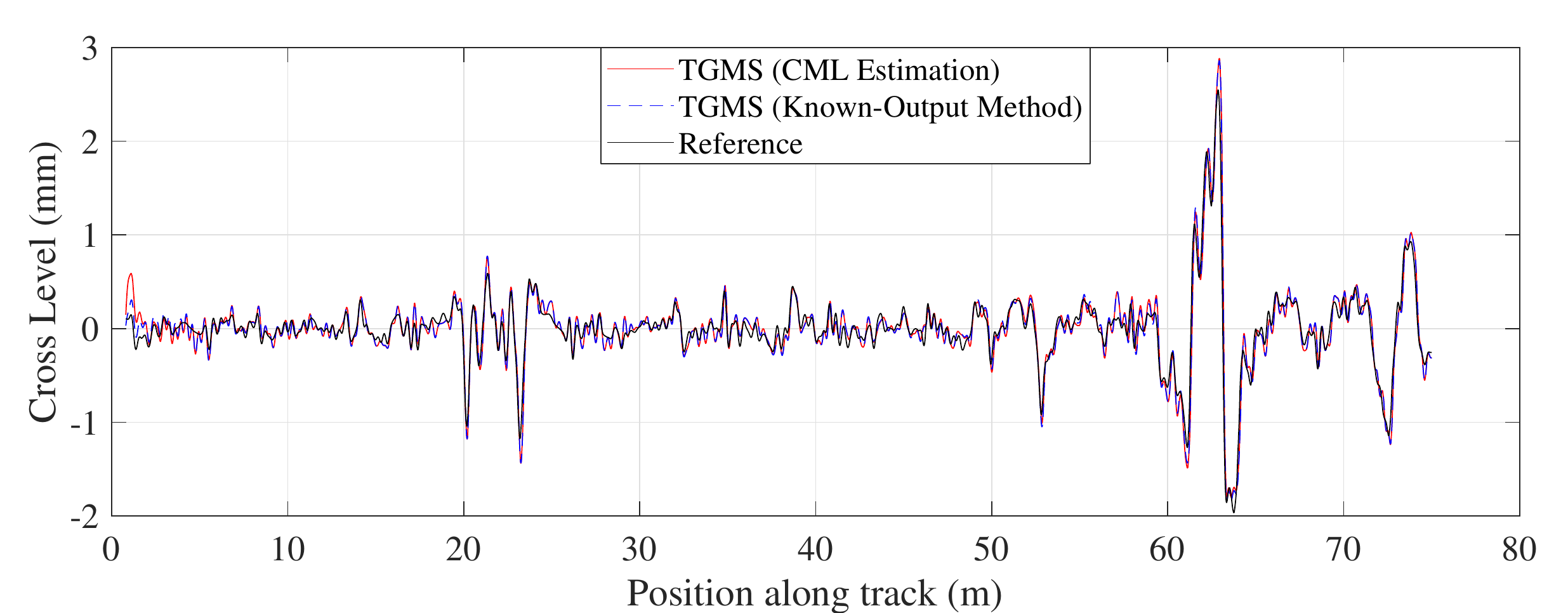}
  \caption{Cross Level obtained with one Kalman filter, compared to the Reference Cross Level. Top: Unfiltered. Bottom: Filtered.}
  \label{fig:Results_CL_1F}
\end{figure}

\begin{figure}[h!]
  \centering
    \includegraphics[width=0.7\textwidth]{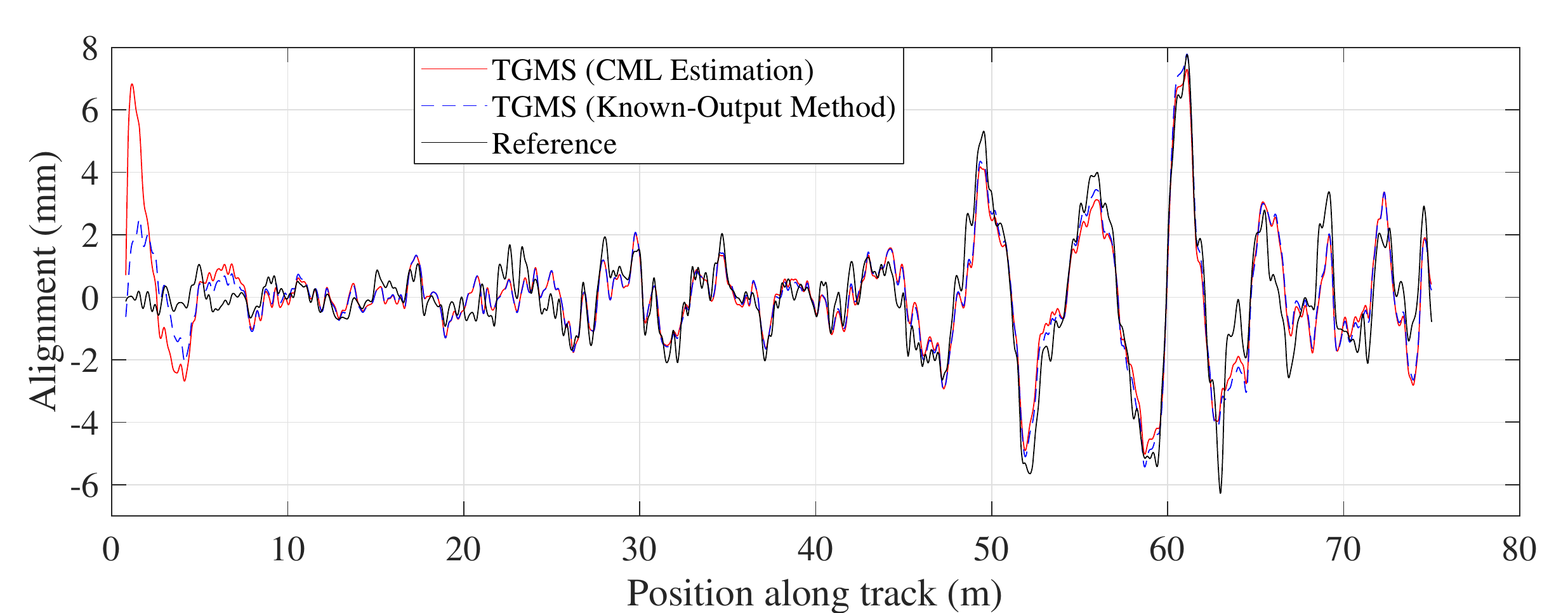}
  \caption{Alignment obtained with one Kalman filter, compared to the Reference Alignment (all signals are filtered).}
  \label{fig:Results_AL_1F}
\end{figure}

\begin{figure}[h!]
  \centering
    \includegraphics[width=0.7\textwidth]{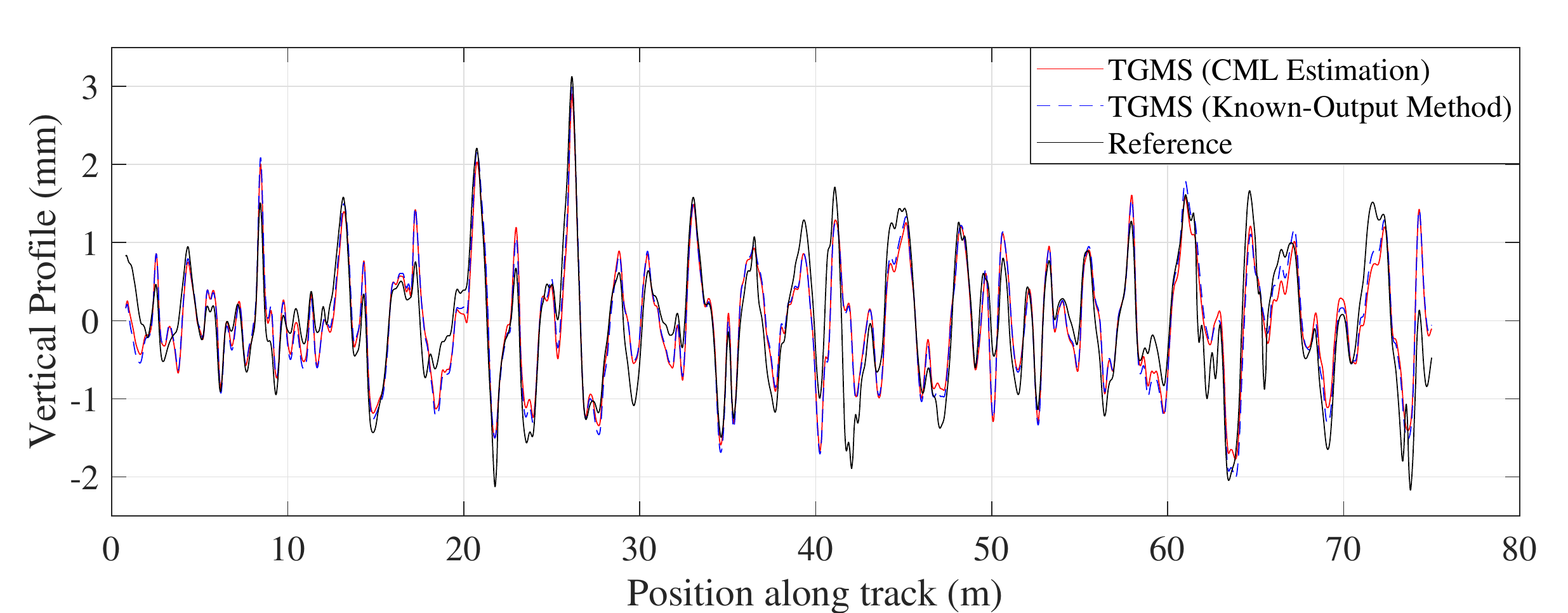}
  \caption{Vertical Profile obtained with one Kalman filter, compared to the Reference Vertical Profile (all signals are filtered).}
  \label{fig:Results_VP_1F}
\end{figure}

Figs.~\ref{fig:Results_GV_1F}-\ref{fig:Results_VP_1F} represent the irregularities obtained when using one Kalman filter for the trajectory and attitude estimation, as described in Section~\ref{sec:FullKalman}. The precision of the CML estimation for covariance matrices (Section~\ref{sec:MLE}) is compared to that of the KOM (Section~\ref{sec:known-output}).

\subsection{Robustness of the known-output method}

It was mentioned in Section~\ref{sec:known-output} that the KOM used to estimate the covariance matrices is based upon the hypothesis that, once the required parameters have been obtained for a specific track, with specific irregularities and vehicle speed, they will remain valid in other conditions. The present subsection intends to test the validity of this assumption.

\begin{figure}[h!]
  \centering
    \includegraphics[width=0.7\textwidth]{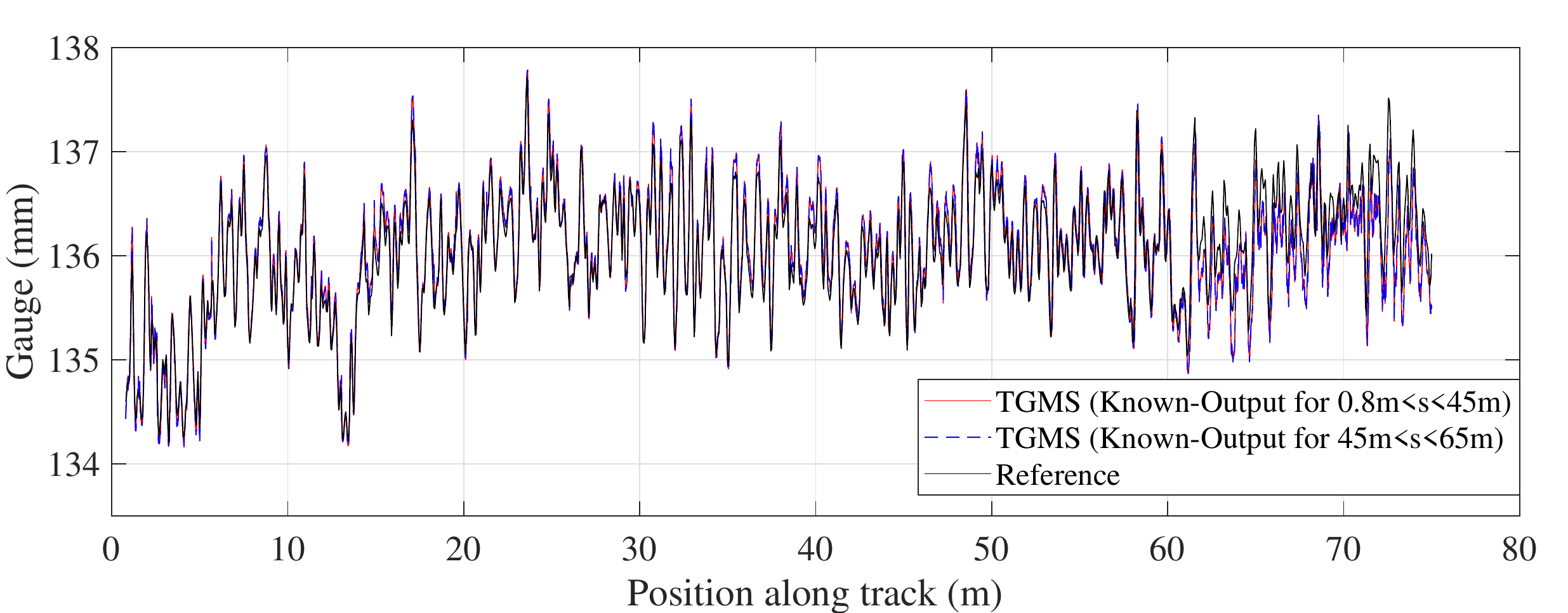}
    \includegraphics[width=0.7\textwidth]{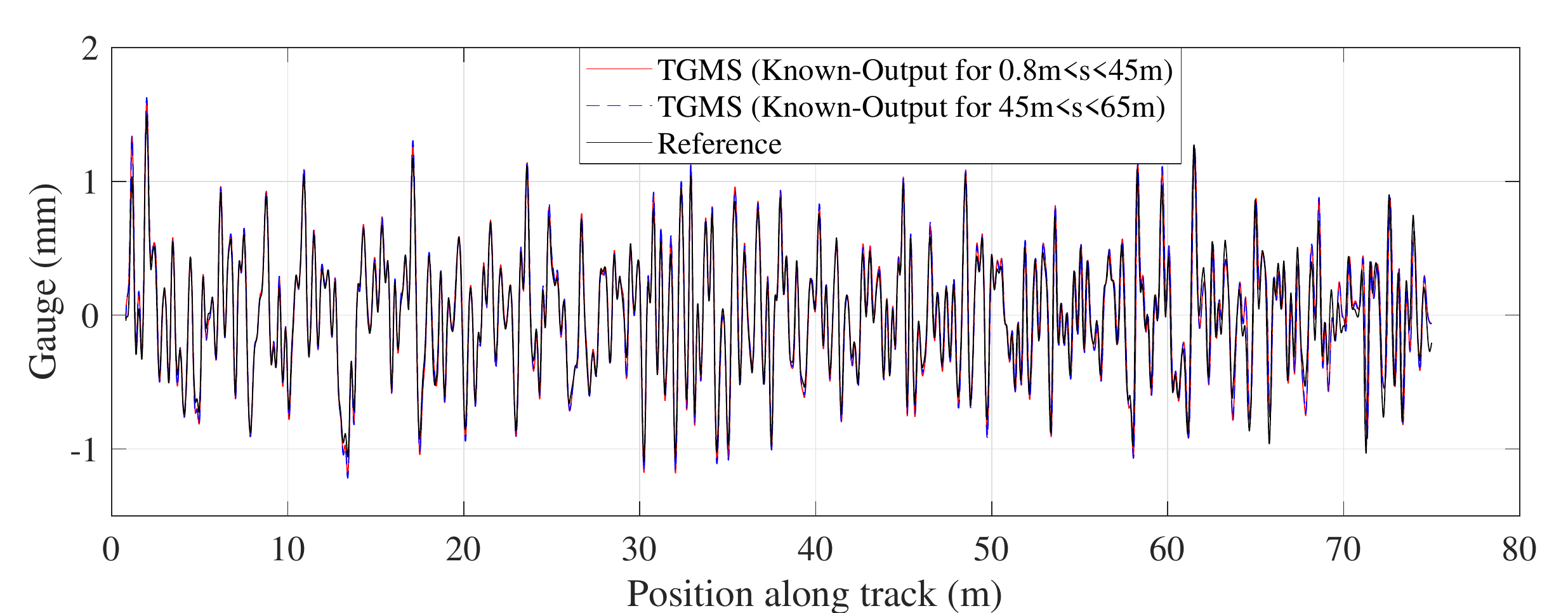}
  \caption {Track Gauge obtained with one Kalman filter and the KOM (optimization conducted for a partial section of the track), compared to the Reference Gauge. Top: Unfiltered. Bottom: Filtered.}
  \label{fig:Results_GV_Rob}
\end{figure}

\begin{figure}[h!]
  \centering
    \includegraphics[width=0.7\textwidth]{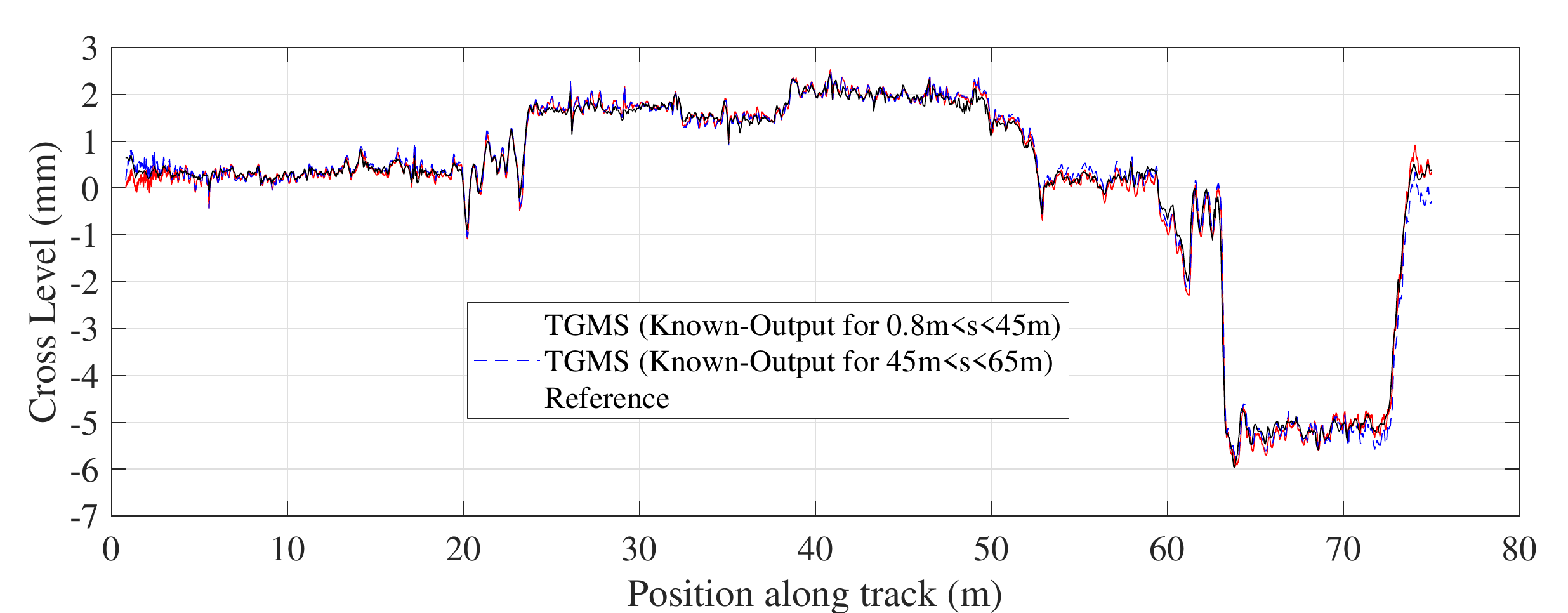}
    \includegraphics[width=0.7\textwidth]{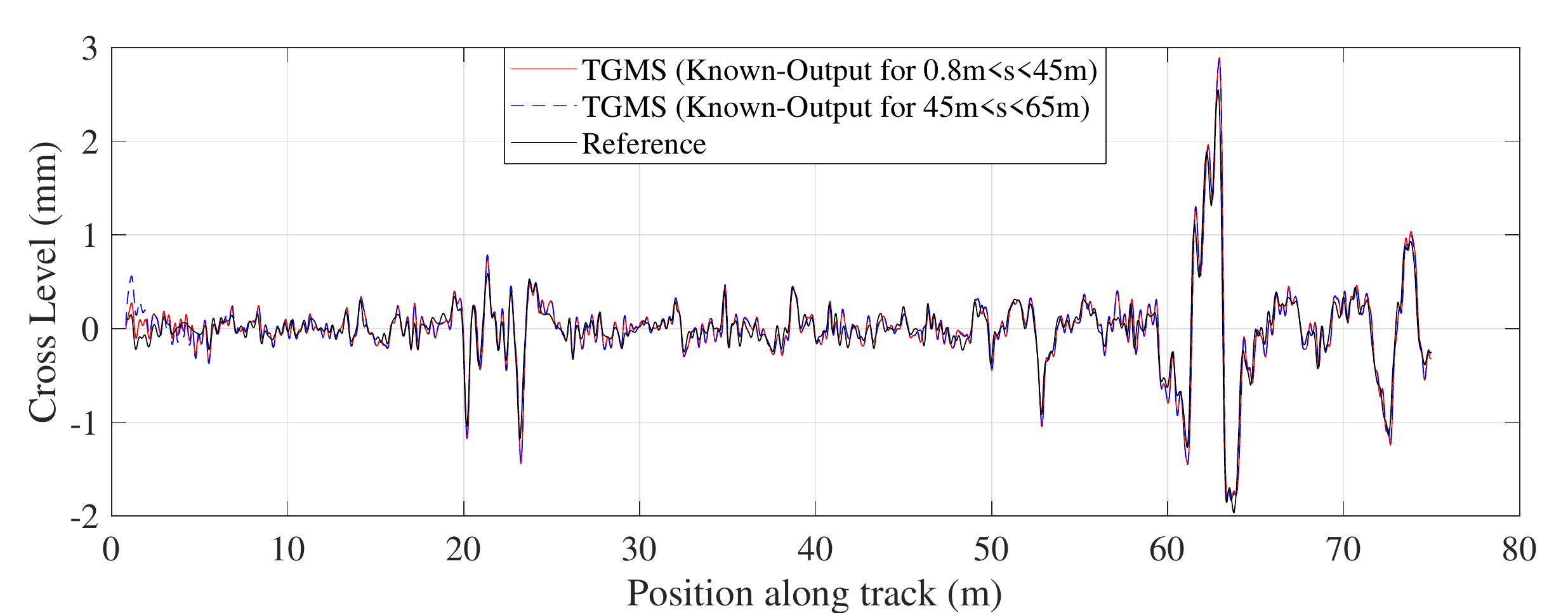}
  \caption{Cross Level obtained with one Kalman filter and the KOM (optimization conducted for a partial section of the track), compared to the Reference Cross Level. Top: Unfiltered. Bottom: Filtered.}
  \label{fig:Results_CL_Rob}
\end{figure}

\begin{figure}[h!]
  \centering
    \includegraphics[width=0.7\textwidth]{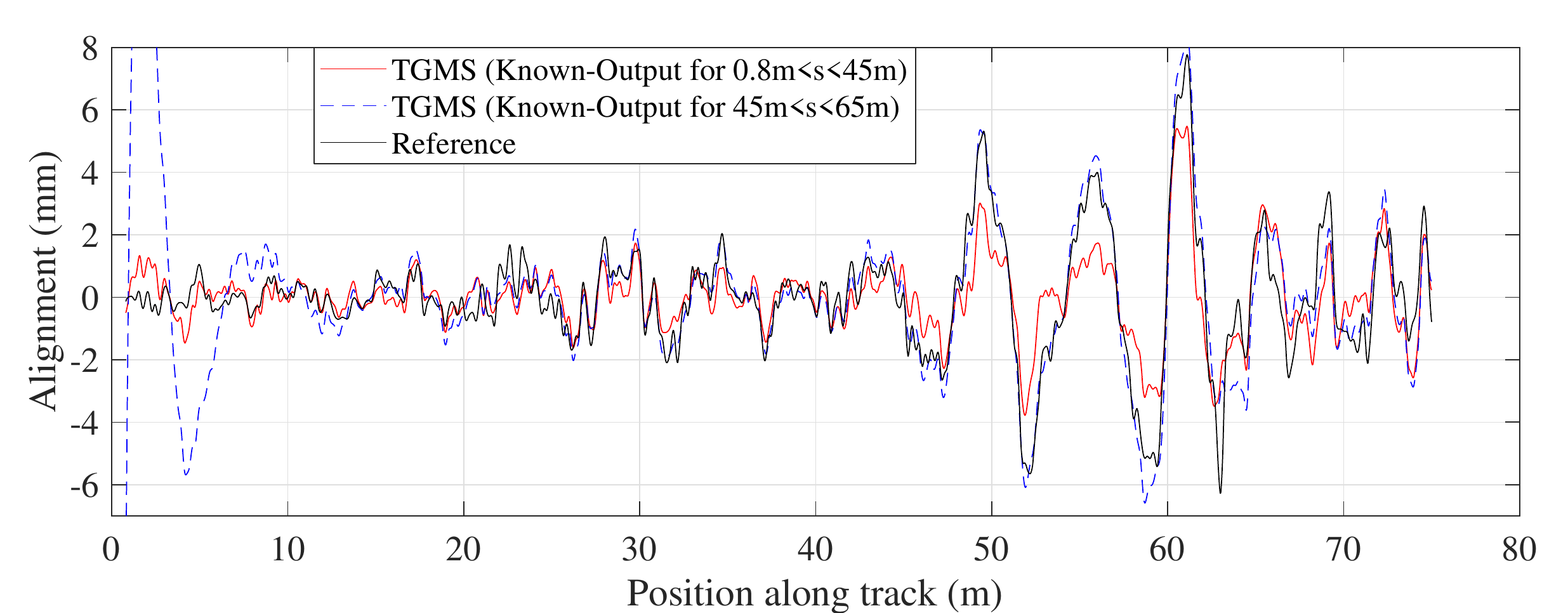}
  \caption{Alignment obtained with one Kalman filter and the KOM (optimization conducted for a partial section of the track), compared to the Reference Alignment (all signals are filtered).}
  \label{fig:Results_AL_Rob}
\end{figure}

\begin{figure}[h!]
  \centering
    \includegraphics[width=0.7\textwidth]{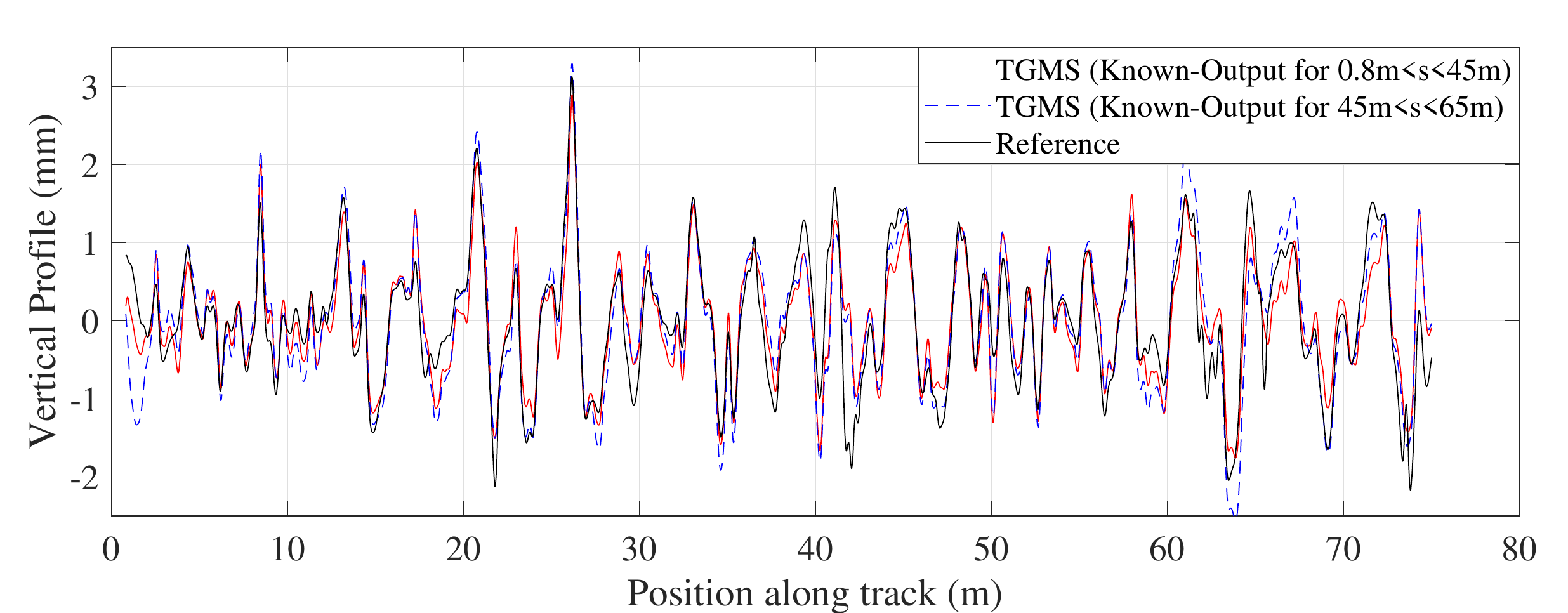}
  \caption{Vertical Profile obtained with one Kalman filter and the KOM (optimization conducted for a partial section of the track), compared to the Reference Vertical Profile (all signals are filtered).}
  \label{fig:Results_VP_Rob}
\end{figure}

One way to verify this would be to to repeat the test with a different distribution of irregularities--recall that the scale track rests upon a set of mechanisms that allow for the generation of any desired profile of irregularities. Although this will be surely done in the near future, a simpler approach that does not require any additional experimental tests has been adopted for this paper. The idea consists in mentally dividing the vehicle ride in 2 parts and checking if the covariance matrices obtained for one of the parts are able to give a good irregularity estimation for the whole test. The results attained, using one Kalman filter, are displayed in Figs.~\ref{fig:Results_GV_Rob}-\ref{fig:Results_VP_Rob}.

\subsection{Discussion}
\label{sec:discussion}

The results shown in Figs.~\ref{fig:Results_GV_2F}-\ref{fig:Results_VP_1F} clearly display a remarkably accurate estimation of the irregularities with all 4 combinations of 1 / 2 Kalman filters, CML estimation / KOM, as evidenced by the similarity between the reference irregularities and those yielded by the TGMS.

As for the robustness of the KOM, evaluated in Figs.~\ref{fig:Results_GV_Rob}-\ref{fig:Results_VP_Rob}, reasonably good results can be found for gauge, cross level and vertical profile. However, the alignment graph does manifest a detrimental effect of estimating the covariance parameters based on only a portion of the track. When optimizing between $\SI{0.8}{\metre}$ and $\SI{45}{\metre}$, the 3 higher alignment peaks are not properly captured, while the optimization between $\SI{45}{\metre}$ and $\SI{65}{\metre}$ produces large errors at the start of the test. These results suggest a limited robustness of the KOM to estimate the covariance matrices. However, if the optimization was carried out over a sufficiently long track with a highly varied frequency content for all irregularities, it might be the case that the obtained covariance parameters yielded good estimations for a wide range of tracks and irregularity levels. In any case, it should be remembered that these issues do not exist when the CML estimation is used, because the algorithm estimates specific covariance matrices for each different ride of the vehicle.

Taking the considerations of the last paragraph into account, it is reasonable to propose the CML estimation for covariance matrices as the most promising algorithm for the implementation of the TGMS technology on commercial vehicles. The KOM may still be useful in order to see how accurate the TGMS results can be when covariance matrices are optimally tuned.

Focusing on the results found with CML estimation, the obtained irregularities are notably precise using either 1 or 2 Kalman filters, with the exception of a spurious alignment peak at the beginning of the test (Figs.~\ref{fig:Results_AL_2F} and \ref{fig:Results_AL_1F}). This initial peak is found to be smaller when using 2 consecutive Kalman filters. On the other hand, a close look to Figs.~\ref{fig:Results_CL_2F} and \ref{fig:Results_CL_1F} reveals that the unfiltered cross level is slightly more accurate with one Kalman Filter. Nevertheless, more experimental results would be needed in order to confirm these small differences. What is apparent from the presented results is that both approaches (CML estimation with 1 Kalman filter / CML estimation with 2 Kalman filters) give rise to sufficiently accurate results for the current application.

Finally, it is appropriate to note that we also tested variants of all presented Kalman filters that include additional state variables to model sensor biases. This was done by assuming the bias variables to evolve in time as Wiener processes, as is common in the literature \cite{keller,anderson}. Although, as expected, this proved to slightly improve the irregularity estimation when using the KOM, the CML estimation strategy was not able to properly tune the parameters associated with the bias variables. For this reason, the results reached by using these augmented Kalman filters are not shown in the paper.

%% file: G_Conclusions.tex
\section{Summary and conclusions} \label{sec:Conclusions}

This paper is about the estimation of the trajectory and attitude of a body moving along a railroad track. Compared with the equivalent, and well developed in the literature, estimation problems used for inertial navigation of air or road vehicles, this estimation problem has two specific properties: (1) the resulting trajectory and attitude are very similar to the known trajectory and attitude of the track centerline, and (2) the required accuracy is very demanding for applications like track geometry measurement. The developed estimation techniques are based on the kinematics of an arbitrary body moving along an arbitrary track and the exact relationship between the description of this type of motion and the measurements of inertial sensors.\\

The estimations are based on the simple discrete Kalman filter. This basic method can be used thanks to an accurate linearization of the kinematics due to the small body to track relative motion. The estimation of trajectory and attitude can be done independently, using a different Kalman filter for each, or as a coupled problem, using a single Kalman filter. Anyway, crucial to the success of the estimation is the accurate calculation of the parameters of the covariance matrices used in the Kalman filters. Two methods are used in this work to calculate these parameters: the KOM and the CML estimation. The KOM, which is specific for the application of track geometry measurement, requires the use of a railroad track with known irregular geometry. The CML estimation is much more general and does not rely upon the previous knowledge of any irregularity profiles. Both techniques are explained in detail.\\

The application of the estimation techniques to the track geometry measurement of a scale track shows the accuracy of the four combinations of techniques that have been presented. In general, results show a remarkably accurate estimation of the track irregularities with all four combinations. Therefore, it is obvious to propose the use of the much simpler and less costly CML method for the calculation of the covariance matrices. This is very good news for the application of the track geometry measurement method proposed by the authors. Regarding the use of one or two Kalman filters, results are not concluding. Thanks to the developed estimation techniques, the accuracy of the measurement of the track irregularities has improved significantly with respect to the results presented in~\cite{escalona2021}.

%% file: H_Appendix.tex
\section*{Appendix A \\
Kalman filter for the orientation: matrices of the state space model} \label{sec:appendixA}

\begin{equation*}
    \textbf{z}=
    \left[
    \begin{array}{*{20}{c}}
          \hat{\omega}_x & \hat{\omega}_y & \hat{\omega}_z & {a_{imu}^{corr}|}_x & {a_{imu}^{corr}|}_y & \psi^t
    \end{array}
    \right]^T
\end{equation*}

\begin{equation*}
\textbf{F}=
\begin{bmatrix}
    \textbf{F}_1 & \textbf{0} & \textbf{0}\\
    \textbf{0} & \textbf{F}_1 & \textbf{0}\\
    \textbf{0} & \textbf{0} & \textbf{F}_1
\end{bmatrix}, \quad \textrm{with} \quad \textbf{F}_1=
\begin{bmatrix}
    1 &  \mathrm{\Delta} t \\
    0 & 1
\end{bmatrix}
\end{equation*}

\begin{equation*}
    \textbf{H}=
    \begin{bmatrix}
    0 & 1 & -\hat{\omega}_z & 0 & 0 & 0 \\
    \hat{\omega}_z & 0 & 0 & 1 & 0 & 0 \\
    -\hat{\omega}_y & 0 & 0 & 0 & 0 & 1 \\
    0 & 0 & -g & 0 & 0 & 0 \\
    g & 0 & 0 & 0 & 0 & 0 \\
    0 & 0 & 0 & 0 & 1 & 0 
    \end{bmatrix}
\end{equation*}

\begin{equation*}
\textbf{Q}=
\begin{bmatrix}
    q_{\varphi}\textbf{Q}_1 & \textbf{0} & \textbf{0}\\
    \textbf{0} & q_{\theta}\textbf{Q}_1 & \textbf{0}\\
    \textbf{0} & \textbf{0} & q_{\psi}\textbf{Q}_1
\end{bmatrix}, \quad \textrm{with} \quad \textbf{Q}_1=
\begin{bmatrix}
    \mathrm{\Delta}t^3/3 &  \mathrm{\Delta}t^2/2 \\
    \mathrm{\Delta}t^2/2 & \mathrm{\Delta}t
\end{bmatrix}
\end{equation*}

\begin{equation*}
\textbf{R}=
\begin{bmatrix}
    R_\omega & 0 & 0 & 0 & 0 & 0 \\
    0 & R_\omega & 0 & 0 & 0 & 0 \\
    0 & 0 & R_\omega & 0 & 0 & 0 \\
    0 & 0 & 0 & R_{ax} & 0 & 0 \\
    0 & 0 & 0 & 0 & R_{ay} & 0 \\
    0 & 0 & 0 & 0 & 0 & R_\psi \\
\end{bmatrix}
\end{equation*}

\section*{Appendix B \\
Kalman filter for the trajectory: matrices of the state space model} \label{sec:appendixB}

\begin{equation*}
    \textbf{z}=
    \begin{bmatrix}
    a_y^{imu} + a_x^{imu}{\psi ^{t,b}} - a_z^{imu}{\varphi ^{t,b}} - g{\varphi ^t} - {\rho _h}{V^2}\\
    a_z^{imu} - a_x^{imu}{\theta ^{t,b}} + a_y^{imu}{\varphi ^{t,b}} - g + {\rho _v}{V^2} \\
    0 \\
    \delta
    \end{bmatrix}
\end{equation*}

\begin{equation*}
\textbf{F}=
\begin{bmatrix}
    \textbf{F}_1 & \textbf{0}\\
    \textbf{0} & \textbf{F}_1\\
\end{bmatrix}, \quad \textrm{with} \quad \textbf{F}_1=
\begin{bmatrix}
    1 &  \mathrm{\Delta}t & \mathrm{\Delta}t^2/2 \\
    0 & 1 & \mathrm{\Delta}t \\
    0 & 0 & 1
\end{bmatrix}
\end{equation*}

\begin{equation*}
\textbf{H}=
\begin{bmatrix}
\textbf{H}_1 & \textbf{H}_2
\end{bmatrix}
\end{equation*}

\begin{equation*}
\textbf{H}_1=
\begin{bmatrix}
-V^2(\rho_{tw}^2+\rho_h^2)& 0 & 1 \\
    \rho_v\rho_hV^2+\rho_{tw}\dot{V} & 2\rho_{tw}V & 0 \\
    1 & 0 & 0 \\
    0 & 0 & 0 
\end{bmatrix}
\end{equation*}

\begin{equation*}
\textbf{H}_2=
\begin{bmatrix}
    \rho_v\rho_hV^2-\rho_{tw}\dot{V} & -2\rho_{tw}V & 0 \\
    -V^2(\rho_{tw}^2+\rho_h^2) & 0 & 1 \\
    0 & 0 & 0 \\
    1 & 0 & 0 
\end{bmatrix}
\end{equation*}

\begin{equation*}
\textbf{Q}=
\begin{bmatrix}
    q_y\textbf{Q}_1 & \textbf{0}\\
    \textbf{0} & q_z\textbf{Q}_1\\
\end{bmatrix}, \quad \textrm{with} \quad \textbf{Q}_1=
\begin{bmatrix}
    \mathrm{\Delta}t^5/20 &  \mathrm{\Delta}t^4/8 & \mathrm{\Delta}t^3/6 \\
    \mathrm{\Delta}t^4/8 & \mathrm{\Delta}t^3/3 & \mathrm{\Delta}t^2/2 \\
    \mathrm{\Delta}t^3/6 & \mathrm{\Delta}t^2/2 & \mathrm{\Delta}t
\end{bmatrix}
\end{equation*}

\begin{equation*}
\textbf{R}=
\begin{bmatrix}
    R_{y1} & 0 & 0 & 0 \\
    0 & R_{z1} & 0 & 0 \\
    0 & 0 & R_{y2} & 0 \\
    0 & 0 & 0 & R_{z2} 
\end{bmatrix}
\end{equation*}

\section*{Appendix C \\
Kalman filter for the orientation and trajectory: matrices of the state space model} 
\label{sec:appendixC}

\begin{equation*}
    \textbf{z}=
    \begin{bmatrix}
    \hat{\omega}_x \\ \hat{\omega}_y \\ \hat{\omega}_z \\
    a_x^{imu}+\theta^t(g-a_z^{imu})+a_y^{imu}\psi^t-\dot{V}\\
    a_y^{imu} + {\varphi ^t}(a_z^{imu}-g) - a_x^{imu}{\psi ^t} - {\rho _h}{V^2}\\
    a_z^{imu} - g - a_y^{imu}{\varphi ^{t}}+ a_x^{imu}{\theta ^{t}}   + {\rho _v}{V^2} \\
    0 \\
    \delta
    \end{bmatrix}
\end{equation*}

\begin{equation*}
\textbf{F}=
\begin{bmatrix}
    \textbf{F}_1 & \textbf{0} & \textbf{0} & \textbf{0} & \textbf{0}\\
    \textbf{0} & \textbf{F}_1 & \textbf{0} & \textbf{0} & \textbf{0}\\
    \textbf{0} & \textbf{0} & \textbf{F}_1 & \textbf{0} & \textbf{0}\\
    \textbf{0} & \textbf{0} & \textbf{0} & \textbf{F}_2 & \textbf{0}\\
    \textbf{0} & \textbf{0} & \textbf{0} & \textbf{0} & \textbf{F}_2\\
\end{bmatrix}
\end{equation*}

\begin{equation*}
\textbf{F}_1=
\begin{bmatrix}
    1 &  \mathrm{\Delta}t \\
    0 & 1 
\end{bmatrix}, \qquad
\textbf{F}_2=
\begin{bmatrix}
    1 &  \mathrm{\Delta}t & \mathrm{\Delta}t^2/2 \\
    0 & 1 & \mathrm{\Delta}t \\
    0 & 0 & 1
\end{bmatrix}
\end{equation*}

\begin{equation*}
\textbf{H}=
\begin{bmatrix}
\textbf{H}_1 & \textbf{H}_2 & \textbf{H}_3
\end{bmatrix}
\end{equation*}

\begin{equation*}
\textbf{H}_1=
\begin{bmatrix}
0 & 1 & -\hat{\omega}_z & 0 & 0 & 0 \\
\hat{\omega}_z & 0 & 0 & 1 & 0 & 0 \\
-\hat{\omega}_y & 0 & 0 & 0 & 0 & 1 \\
0 & 0 & -a_z^{imu} & 0 & a_y^{imu} & 0 \\
a_z^{imu} & 0 & 0 & 0 & -a_x^{imu} & 0 \\
-a_y^{imu} & 0 & a_x^{imu} & 0 & 0 & 0 \\
0 & 0 & 0 & 0 & 0 & 0 \\
0 & 0 & 0 & 0 & 0 & 0 \\
\end{bmatrix}
\end{equation*}

\begin{equation*}
\textbf{H}_2=
\begin{bmatrix}
    0 & 0 & 0 \\
    0 & 0 & 0 \\
    0 & 0 & 0 \\
    V^2(\rho_{tw}\rho_v-\rho'_h)-\rho_h\dot{V}  & -2\rho_hV & 0\\
    -V^2(\rho_{tw}^2+\rho_h^2)& 0 & 1 \\
    \rho_v\rho_hV^2+\rho_{tw}\dot{V} & 2\rho_{tw}V & 0 \\
    1 & 0 & 0 \\
    0 & 0 & 0 
\end{bmatrix}
\end{equation*}

\begin{equation*}
\textbf{H}_3=
\begin{bmatrix}
    0 & 0 & 0 \\
    0 & 0 & 0 \\
    0 & 0 & 0 \\
    \rho_v\dot{V}+\rho_{tw}\rho_hV^2 & 2\rho_vV & 0\\
    \rho_v\rho_hV^2-\rho_{tw}\dot{V} & -2\rho_{tw}V & 0 \\
    -V^2(\rho_{tw}^2+\rho_h^2) & 0 & 1 \\
    0 & 0 & 0 \\
    1 & 0 & 0 
\end{bmatrix}
\end{equation*}

\begin{equation*}
\textbf{Q}=
\begin{bmatrix}
    q_\varphi\textbf{Q}_1 & \textbf{0} & \textbf{0} & \textbf{0} & \textbf{0}\\
    \textbf{0} & q_\theta\textbf{Q}_1 & \textbf{0} & \textbf{0} & \textbf{0}\\
    \textbf{0} & \textbf{0} & q_\psi\textbf{Q}_1 & \textbf{0} & \textbf{0}\\
    \textbf{0} & \textbf{0} & \textbf{0} & q_y\textbf{Q}_2 & \textbf{0}\\
    \textbf{0} & \textbf{0} & \textbf{0} & \textbf{0} & q_z\textbf{Q}_2
\end{bmatrix}
\end{equation*}

\begin{equation*}
\textbf{Q}_1=
\begin{bmatrix}
    \mathrm{\Delta}t^3/3 & \mathrm{\Delta}t^2/2 \\
    \mathrm{\Delta}t^2/2 & \mathrm{\Delta}t
\end{bmatrix}, \qquad 
\textbf{Q}_2=
\begin{bmatrix}
    \mathrm{\Delta}t^5/20 &  \mathrm{\Delta}t^4/8 & \mathrm{\Delta}t^3/6 \\
    \mathrm{\Delta}t^4/8 & \mathrm{\Delta}t^3/3 & \mathrm{\Delta}t^2/2 \\
    \mathrm{\Delta}t^3/6 & \mathrm{\Delta}t^2/2 & \mathrm{\Delta}t
\end{bmatrix}
\end{equation*}

\begin{equation*}
\textbf{R}=
\begin{bmatrix}
    R_\omega & 0 & 0 & 0 & 0 & 0 & 0 & 0 \\
    0 & R_\omega & 0 & 0 & 0 & 0 & 0 & 0 \\
    0 & 0 & R_\omega & 0 & 0 & 0 & 0 & 0 \\
    0 & 0 & 0 & R_x & 0 & 0 & 0 & 0 \\
    0 & 0 & 0 & 0 & R_{y1} & 0 & 0 & 0 \\
    0 & 0 & 0 & 0 & 0 & R_{z1} & 0 & 0 \\
    0 & 0 & 0 & 0 & 0 & 0 & R_{y2} & 0 \\
    0 & 0 & 0 & 0 & 0 & 0 & 0 & R_{z2}
\end{bmatrix}
\end{equation*}